\documentclass[preprint,showpacs,preprintnumbers,onecolumn,superscriptaddress,floatfix,12pt]{revtex4-1}
\usepackage[utf8]{inputenc}
\usepackage{bm,amsmath,amssymb}
\usepackage{graphicx}
\usepackage{color}
\usepackage{hyperref}
\usepackage{multirow}
\usepackage[compatibility=false]{caption}
\usepackage{subcaption}
\usepackage{soul}
\usepackage{float}
\usepackage[normalem]{ulem}

\let\oldalign\align
\def\align{\linenomath\oldalign}

\begin{document}

\title{The spatial string tension and its effects on screening correlators in a thermal QCD plasma }

\author{Dibyendu Bala}
\affiliation{Fakult\"{a}t f\"{u}r Physik, Universit\"{a}t Bielefeld, D-33615 Bielefeld, Germany}
\author{Olaf Kaczmarek}
\affiliation{Fakult\"{a}t f\"{u}r Physik, Universit\"{a}t Bielefeld, D-33615 Bielefeld, Germany}
\author{Peter Petreczky}
\affiliation{Physics Department, Brookhaven National Laboratory, Upton, New York 11973, USA}
\author{Sayantan Sharma}
\affiliation{The Institute of Mathematical Sciences, a CI of Homi Bhabha National Institute, Chennai, 600113, India}
\author{Swagatam Tah}
\email{Corresponding author: swagatamt@imsc.res.in}
\affiliation{The Institute of Mathematical Sciences, a CI of Homi Bhabha National Institute, Chennai, 600113, India}

\begin{abstract}

We calculate the spatial Wilson line correlator for $2+1$ flavor QCD using 
highly improved staggered quark discretization for fermions and in quenched 
QCD for a wide range of temperatures,  from the chiral crossover 
temperature $\mathrm{T_{pc}\simeq 156}$ MeV or the deconfinement temperature $\simeq 300$ 
MeV respectively, up to $2$ GeV. Extracting the spatial string tension for different 
lattice cut-offs and by performing a continuum extrapolation of this observable, we show 
that the soft (magnetic) gluons interact non-perturbatively even at temperatures $\gtrsim 1$ GeV. 
We provide incriminating evidences to demonstrate that dimensionally reduced effective theories 
can describe these soft quark and gluon quasi-particles for both quenched 
and $2+1$ flavor QCD, at temperatures  $\mathrm{T\gtrsim 5T_{pc}}$. We also show for 
the first time the imprints of the non-perturbative pseudo-potential 
in the properties of mesonic screening masses for temperatures ranging from 
$0.8$-$164$ GeV in the quark-gluon plasma. 
\end{abstract}

\maketitle


 \textbf{Introduction:}
Quantum Chromodynamics (QCD) has a rich set of non-perturbative 
properties at finite temperatures~\cite{Gross:1980br,Shuryak:1993kg}. There  is a transition from a phase of 
color-singlet degrees of freedom to a phase where color charges are  deconfined when the temperature is increased, 
which occur through a first order phase transition~\cite{Boyd:1995zg} or a smooth crossover~\cite{Aoki:2006we} 
respectively in QCD without and in the presence of dynamical quarks.  Ab-initio lattice field 
theory calculations have demonstrated this and also determined the corresponding transition temperature which 
is $\mathrm{T_d r_0= 0.7457(45)}$~\cite{Francis:2015lha} for QCD without quarks (pure SU(3) gauge theory or quenched QCD), 
whereas it is $\mathrm{T_{pc} \simeq 156}$ MeV~\cite{HotQCD:2018pds, Burger:2018fvb, Borsanyi:2020fev, Gavai:2024mcj} 
in presence of two degenerate light  quarks and one strange quark flavor. In the deconfined phase of QCD 
 with a gauge coupling  $\mathrm{g}$, two more scales apart from the temperature $\mathrm{T}$ are dynamically generated; 
 the electric $\mathrm{gT}$ and the magnetic scale $\mathrm{g^2T/\pi}$, which correspond to the inverse correlation 
 lengths of the color-electric and color-magnetic degrees of freedom respectively. When these scales are well separated 
 i.e. $\mathrm{g^2 T/\pi \ll gT \ll \pi T}$, the gauge and fermion fields with typical momenta $\mathrm{\sim \pi T}$ 
 can be integrated out~\cite{Appelquist:1974tg} resulting in dimensionally reduced effective theory known 
 as electro-static QCD (EQCD) ~\cite{Nadkarni:1982kb,Braaten:1995jr}. Within this effective field theory the 
 temporal component of the gauge field $\mathrm{A_0}$ appear as a static adjoint scalar field and the parameters 
 of this theory i.e. the couplings $\mathrm{g_E, \lambda_E}$ and the Debye mass $\mathrm{m_D}$ can be calculated 
 using perturbation theory to all orders in $\mathrm{g}$~\cite{Braaten:1995jr,Kajantie:1997tt}. The high temperature phase 
 in QCD  correspond to the confined phase in a 3 dimensional adjoint Higgs theory\cite{Kajantie:1997tt,Karsch:1998tx}. The 
 confinement scale of this theory corresponds to the magnetic scale in QCD, which is inherently non-perturbative and gives 
 rise to infrared divergences which  cannot be tamed through perturbative resummations~\cite{Linde:1980ts}.  At higher 
 temperatures, the $\mathrm{A_0}$ fields  become massive and can be integrated out from the QCD partition function, 
 giving rise to an effective theory called magneto-static QCD (MQCD) which is a 3 dimensional pure gauge 
 theory~\cite{Appelquist:1981vg,DHoker:1980rnd,Braaten:1995jr}.  However it is not clear a priori at what 
 temperatures such a clean separation  of scales would occur to justify the validity of these effective 
 theories to describe QCD~\cite{Laine:1999hh,Gavai:2000mx}.

The non-perturbative magnetic sector in high temperature QCD can be directly accessed through the 
calculation of the spatial Wilson loop~\cite{Svetitsky:1982gs,Borgs:1985qh}. Earlier lattice calculations 
have demonstrated the area-law behavior of the spatial Wilson loop, further extracting the spatial string 
tension $\mathrm{\sigma_s}$ from the linearly rising pseudo-potential at large distance scales in SU(3) without 
dynamical quarks~\cite{Manousakis:1986jh,Karkkainen:1993ch,Bali:1993tz,Karsch:1994af,Boyd:1996bx} 
as well as in $2+1$ flavor QCD with a heavier than physical light quark mass \cite{Cheng:2008bs}. 
An agreement with the estimate of $\mathrm{\sigma_s}$ in dimensional reduced theory was observed at 
$\mathrm{\gtrsim 2~T_{pc}}$, and agreement between the EQCD calculations and lattice QCD data of different 
thermodynamic quantities like the quark number susceptibilities are typically observed at temperatures 
$\mathrm{\gtrsim 3~ T_{pc}}$~\cite{Haque:2024gva}.

In this Letter, we revisit the calculation of the $\mathrm{\sigma_s}$ using sophisticated lattice QCD techniques 
to convincingly address the question \emph{what temperature range} herald the onset of the dimensional 
reduction in QCD ?  By performing a careful continuum extrapolation of the pseudo-potential 
extracted out of spatial Wilson line correlators calculated at two(three) different lattice cut-offs for a 
temperature range $\sim$ 0.2-1(2.7) GeV for QCD with(without) dynamical physical quarks, we demonstrate 
the onset of dimensional reduction at temperatures $\mathrm{\gtrsim 5 ~T_{pc}}$ from a detailed understanding 
of its properties. 
We show for the first time that this non-perturbative potential describing the magnetic modes can
remarkably explain the deviation of the screening masses of meson-like long-distance excitations at high 
temperatures $0.8$-$164$ GeV from the perturbative EQCD prediction. We further quantify the spin-splitting 
between the vector and the pseudo-scalar screening excitations which cannot be addressed within the leading 
order perturbative EQCD framework~\cite{Laine:2003bd}.

\textbf{Theoretical and numerical framework:}
In this work we extract the \emph{confining} pseudo-potential and subsequently $\mathrm{\sigma_s}$ by measuring 
the correlation between two spatial Wilson lines $\mathrm{W(R,L)}$ in modified-Coulomb gauge, each of which are 
of length $\mathrm{L}$ along the z direction separated by a transverse distance $\mathrm{R}$ in the x-y plane. 
We use spatial Wilson line correlators in modified-Coulomb gauge instead of the spatial Wilson loops because 
these have a much better signal to noise ratio and lead to the same pseudo-potential. This approach is often 
used to obtain the static potential in zero temperature 
QCD~\cite{Aubin:2004wf,Bazavov:2011nk,HotQCD:2014kol,Bazavov:2017dsy,Brambilla:2022het},
and results in the same potential as the calculations using Wilson loops, see e.g.~\cite{Cheng:2007jq}.
We implement the Wilson lines on a 4 dimensional Euclidean lattice with $\mathrm{N}$ sites along each spatial 
direction and $\mathrm{N_\tau}$ sites along the temporal direction, $\tau$ and at each time-slice. We consider 
$\mathrm{N}=32, 40$ for 2+1 flavor QCD and $\mathrm{N}=32, 48, 64$ for pure SU(3) with $\mathrm{N_\tau=N/4}$ for 
each temperature $\mathrm{T=1/(N_\tau.a)}$ which is varied between $0.16$-$1$ GeV and $0.3$-$2.7$ GeV respectively. 
The pseudo-potential $\mathrm{V(R)}$ can be extracted from the ratio of expectation values of the spatial Wilson 
line correlators of length $\mathrm{z=L.a}$ and $\mathrm{z=(L+1).a}$ where $\mathrm{L \lesssim N}$ from the relation,
\begin{equation}
    \mathrm{V(R).a = \lim_{L\rightarrow \infty} \ln \left[{\frac{ \langle W(R,L) \rangle}{\langle W(R,L+1) \rangle}}\right]}~.
    \label{eqn:potentialdef}
\end{equation}

\begin{figure}[H]
    \centering
    \includegraphics[width=0.8\textwidth]{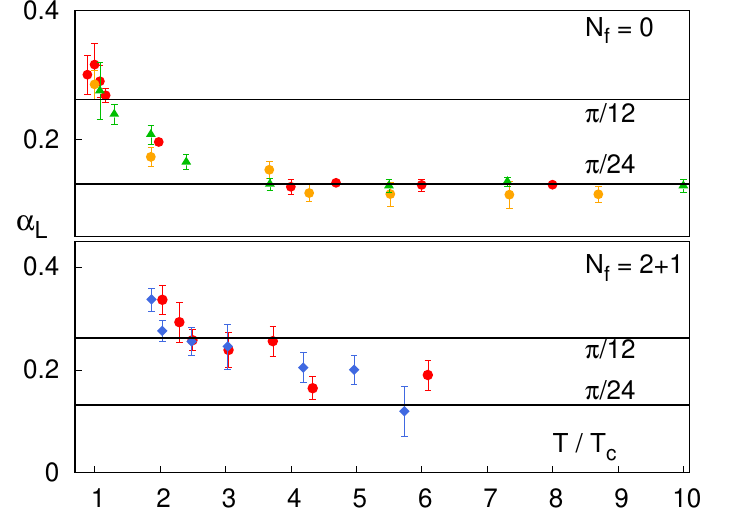}
    \caption{ The coefficient of the Coulomb term extracted from the fit to $\mathrm{V(r)}$ at 
    intermediate distances $\mathrm{(\pi T)^{-1} < r <  (g^2T/\pi)^{-1}}$ for $\mathrm{N_f=0}$ and 
    $\mathrm{N_f=2+1}$ flavor QCD for a     range of temperatures between $1$-$\mathrm{10~T_c}$. 
    The red, blue, orange and green points correspond to lattice size $\mathrm{N_\tau = 8, 10, 12, 16}$ 
    respectively.}
\label{fig:Luscher term Pure gauge and 2+1 flavor QCD}
\end{figure} 

The expectation values are calculated typically over 200-2400 gauge configurations 
in QCD with and without dynamical fermions, details of which are mentioned in the supplementary 
material.  Due to finite statistics the right side of Eq.~\ref{eqn:potentialdef} is dominated 
by the gauge noise for large distances $\mathrm{L > 5}$-$10$ due to which a clear plateau-like 
feature is not visible in this observable. 
This is due to finite lattice size as well as contamination from the excited states. Hence
we use an ansatz $\mathrm{m(L,R)~=~V(R)+b\cdot \rm{e}^{-V_1\cdot L}}$ in order to 
extract the ground state energy $\mathrm{V(R)}$ reliably at each $\mathrm{R}$ by removing the contamination 
due to the first excited state energy $\mathrm{V_1}$. We have varied the fit range between $\mathrm{L \in [1,4]}$ 
for $\mathrm{T\lesssim 4~T_c}$ and $\mathrm{L\in [1,8]}$ at $\mathrm{T\gtrsim 4 ~T_c}$. The pseudo-potential 
is extracted for each jack-knife bin and the statistical errors are estimated from the variance 
among such bins. For the highest temperatures we have studied, the spatial volumes are small leading 
to contamination from the excited states leading to larger errors in the extracted 
pseudo-potential. 

The pseudo-potential $\mathrm{V(R)}$ has an additional divergence in the continuum 
due to the self energy of quark and anti-quark pair. We have renormalized $\mathrm{V(R)}$ using 
the same renormalization factor 
$(\mathrm{C_Q})$~\cite{Bazavov:2016uvm} which is extracted 
from the static potential between a quark-antiquark pair at 
$\mathrm{T=0}$. This is allowed since no new ultraviolet 
divergences appear at finite temperatures. Further details on 
our procedure to extract the pseudo-potential and the results 
for pseudo-potentials at different temperatures and 
lattice sizes are shown in 
Fig.~\ref{fig:PaperPseudoPotentialNt8Nt10_2+1flavourQCD} and 
discussed in the supplementary material.

\textbf{Spatial string tension \& the onset of dimensional reduction:}
Once the renormalized pseudo-potential is measured, we extract the spatial string 
tension using the ansatz $\mathrm{V(r) = V_0 + \sigma_s r - \alpha_c/r}$. We also extract the 
coupling $\mathrm{\alpha_c}$ as a result of the fit.
The  Coulomb term in the above ansatz could have two possible origins.
It could come from  the short distance perturbative part, in which case we have 
$\mathrm{\alpha_c}=\mathrm{\alpha_P}=\mathrm{(4/3) g^2/(4 \pi)}$ at leading order.
Alternatively, the Coulomb term can arise from the fluctuations of the 
color flux tube along the transverse direction~\cite{Luscher:1980ac,Luscher:2002qv,Luscher:1980fr,Alvarez:1981kc}, 
in which case $\mathrm{\alpha_c}=\mathrm{\alpha_L~=~(D-2)\frac{\pi}{24}}$,
with $\mathrm{D}$ being the spacetime dimension. This is the so-called L\"uscher term 
and the above description applies at relatively large distances.  When 
dimensional reduction sets in $\mathrm{D=3}$, otherwise we have $\mathrm{D=4}$.
One would expect that for $\mathrm{r < (\pi T)^{-1}}$ 
the perturbative $\mathrm{\alpha_P/r}$ term will dominate the pesudo-potential, whereas for intermediate 
distances $\mathrm{(\pi T)^{-1} < r < (g^2 T/\pi)^{-1}}$ the L\"{u}scher term determined by $\mathrm{\alpha_L}$ will 
be relevant. At large distances $\mathrm{r> (g^2 T/\pi)^{-1}}$, one would expect the largest contribution 
coming from the term containing $\mathrm{\sigma_s}$. Indeed we observe that the $\mathrm{\alpha_c}$ extracted 
from our fits are close to $\mathrm{\alpha_L}$ at intermediate distances $\mathrm{r.T \in [0.5,1.2]}$  
and eventually tending towards zero at larger distances. 

\begin{figure}[H]
    \centering
    \includegraphics[width=0.7\textwidth]{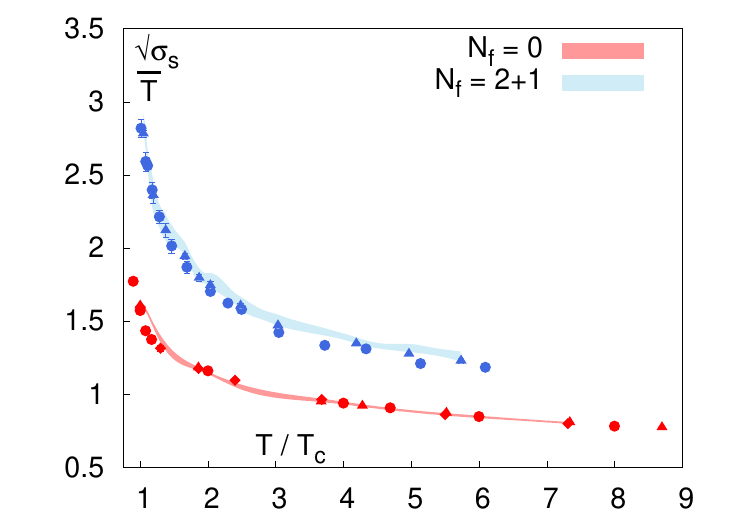}
    \caption{ The comparison of the continuum extrapolated values of $\mathrm{\sqrt{\sigma_s}/T}$ as a function of $\mathrm{T/T_c}$ for quenched (red band) and $2+1$ flavor QCD (blue band). The lattice data for cutoff corresponding to $\mathrm{N_\tau = 8,10,12,16}$ are shown as circle, triangle, box and diamond symbols respectively.}
    \label{fig:StringtensionQcdvsPureGauge}
\end{figure}

The results for the extracted $\mathrm{\alpha_L}$ is compiled in 
Fig.~\ref{fig:Luscher term Pure gauge and 2+1 flavor QCD}. 
The average values of $\mathrm{\alpha_L}$ are consistent with 
$\mathrm{\pi/12}$ for $\mathrm{T \lesssim 5~T_c~(2~T_c)}$ but 
beyond $\mathrm{T \gtrsim 5~T_c~(3~T_c)}$, these values 
approach $\mathrm{\pi/24}$ indicating the onset 
of dimensional reduction beyond this temperature for QCD 
with (without) dynamical quarks. Henceforth $\mathrm{T_c}$ 
correspond to $\mathrm{T_{pc}~(T_d)}$ for SU(3) with 
(without) dynamical quarks. To ensure the robustness of our extraction 
method for $\mathrm{\alpha_L}$ we have compared our results with a 
previous lattice study~\cite{Boyd:1996bx} in SU(3) gauge theory with $\mathrm{N_f=0}$. 
We observe a good agreement with the earlier results. However going to finer 
$\mathrm{N_\tau=12,16}$ lattice spacings is important to precisely demonstrate 
that the value of $\mathrm{\alpha_L}$ approaches $\mathrm{ \pi/24}$ 
at temperatures $\mathrm{T\gtrsim4~T_c}$.
From these observations, we thus held the coefficient of $\mathrm{1/r}$ term to a fixed
value depending upon the temperature and extract $\mathrm{\sqrt{\sigma_s}/T}$ 
from a two-parameter fit to the pseudo-potential for SU(3) gauge 
theory with and without dynamical quarks as a function of 
$\mathrm{T/T_c}$, results of which are shown in 
Fig.~\ref{fig:StringtensionQcdvsPureGauge}. 
We observe that $\mathrm{\sqrt{\sigma_s}/T}$ has a strong temperature 
dependence for $\mathrm{T<4~T_c}$, beyond which it is mildly
dependent of temperature for $\mathrm{N_f=0}$ case, whereas for 
$\mathrm{N_f=2+1}$, a slow variation with $\mathrm{T/T_c}$ persist 
till $\mathrm{\sim 5~T_c}$. This coincides with the onset of dimensional 
reduction and the 3 dimensional effective theory
being confining with the scale characterized by $\mathrm{\sqrt{\sigma_s}}$. 

\begin{figure}[H]
    \centering
    \includegraphics[width=0.7\textwidth]{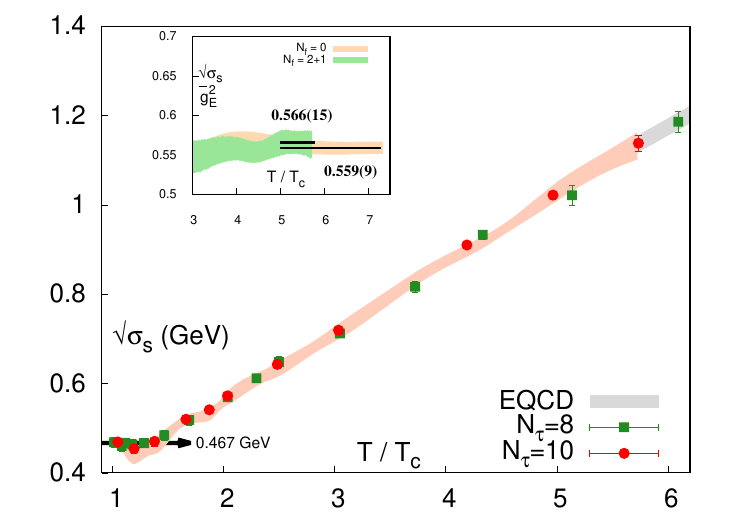}
    \caption{The variation of $\mathrm{\sqrt{\sigma_s}}$ with $\mathrm{T/T_c}$ for different 
    lattice cut-offs corresponding to $\mathrm{N_\tau=8,10}$ along with its continuum estimates (orange 
    band). The continuum estimates are close to EQCD prediction at $\mathrm{T\gtrsim 5~T_c}$, shown 
    as gray band. The black line in lower left corner denotes $\mathrm{\sqrt \sigma=0.467}$ GeV at
    $\mathrm{T=0}$ from Ref.~\cite{Brambilla:2022het}. The variation of the continuum estimate for 
    $\mathrm{\sqrt{\sigma_s}/g_E^2}$ with $\mathrm{T/T_c}$ is shown in the inset for SU(3) (yellow band) 
    and $2+1$ flavor QCD (green band) respectively.}
    \label{fig:String_tensionQCD}
\end{figure}

We next interpret the temperature dependence of the spatial string tension 
in QCD. The extracted spatial string tension for $2+1$ flavor QCD, for different lattice 
spacings or equivalently different $\mathrm{N_\tau=8,10}$ for the entire temperature range 
we have studied, is shown in physical units in Fig.~\ref{fig:String_tensionQCD}. The data 
for $\mathrm{\sqrt{\sigma_s}}$ do not show a significant lattice cut-off dependence as evident 
from the good agreement between $\mathrm{N_\tau=8}$ and $\mathrm{N_\tau=10}$ data. Nevertheless 
we perform a cubic splines interpolation of the data as a function of temperature and for each 
$\mathrm{N_\tau}$ and then perform a continuum extrapolation, which is shown as a red band in 
Fig.~\ref{fig:String_tensionQCD}. We observe that at $\mathrm{T<260}$ MeV the value of 
$\mathrm{\sqrt{\sigma_s}}$ is in excellent agreement with the string tension, 
$\mathrm{\sqrt{\sigma}=0.467}$ GeV extracted from the temporal Wilson line correlator~\cite{Brambilla:2022het} 
at $\mathrm{T=0}$~\cite{Manousakis:1986jh}. At temperatures $\gtrsim 260$ MeV the electric scale $\mathrm{gT}$ 
starts to separate from the hard scale $\mathrm{\pi T}$ resulting in the onset of linear temperature dependence 
of $\mathrm{\sqrt{\sigma_s}}$.

At higher temperatures $\mathrm{\sqrt {\sigma_s}}$ rises with increasing temperatures and we would 
like to quantify this trend. To compare with the literature, the variation of $\mathrm{\sqrt{\sigma_s}/g_E^2}$ with 
temperature is shown in the inset of Fig.~\ref{fig:String_tensionQCD}. The 3 dimensional EQCD coupling 
$\mathrm{g_E^2}$ is calculated up to two-loop in $\mathrm{g^2}$~\cite{Laine:2005ai}. It is evident from the plot that 
the dependence of $\mathrm{\sqrt{\sigma_s}/g_E^2}$ on temperature (or equivalently the 4 dimensional 
gauge coupling $\mathrm{g}$) is mild for $\mathrm{T > 5~ T_c}$, again reinforcing the validity of an 
effective 3 dimensional description as also observed in the data for the L\"uscher term.  The other couplings within EQCD 
i.e. $\mathrm{m_D,\lambda_E}$ have an even milder dependence on $\mathrm{g}$~\cite{Laine:2005ai} hence fitting 
the data for $\mathrm{\sqrt{\sigma_s}/g_E^2}$ with a constant function for the range of temperatures between 
$5$-$\mathrm{5.8~T_c}$, we obtain $\mathrm{\sqrt{\sigma_s}=0.566(15)~g_E^2}$. Extrapolating our 
EQCD inspired fit to higher temperatures, shown as a gray band in Fig.~\ref{fig:String_tensionQCD}, we observe 
that it perfectly accommodates for the lattice data at a cut-off corresponding to $\mathrm{N_\tau=8, T/T_c\sim 6}$, 
which was not used in the fit.  Our results are consistent with the first extensive study in 2+1 flavor QCD performed 
with a different staggered fermion discretization~\cite{Cheng:2008bs}, where lattice data for $\mathrm{\sqrt{\sigma_s}}$ 
was found to match with $\mathrm{0.54(1)g_E^2}$, albeit with a large uncertainty, at temperatures 
$\mathrm{T \gtrsim 1.5~T_c}$ where $\mathrm{T_c\sim 200}$ MeV. As already discussed in Ref.~\cite{Cheng:2008bs} 
major sources of uncertainties in such a direct comparison with the prediction from dimensionally reduced 
theory are due to errors in the determination of spatial string tension and the coupling $\mathrm{g_E}$ in 
a 3 dimensional SU(3)-Higgs model. We have instead changed the strategy and extracted the dependence of the 
spatial string tension with $\mathrm{g_E^2}$ with a constant fit to our data for $\mathrm{\sqrt{\sigma_s}/g_E^2}$ 
in a temperature range where we observe hints of the onset of dimensional reduction. 

We have also performed the same analysis for SU(3) gauge theory without dynamical quarks but for a 
 wider temperature range, which allows us to observe the onset of dimensional reduction more carefully. Since our 
 continuum extrapolated data for $\mathrm{\sqrt{\sigma_s}/g_E^2}$ in SU(3) shows almost no temperature 
 or equivalently the 4 dimensional gauge coupling dependence  in the range $\mathrm{5<T/T_c< 7.3}$, hence 
 it can be  parametrized as $\mathrm {0.559(9) g_E^2}$ in this range and beyond as shown in the inset 
 in Fig.~\ref{fig:String_tensionQCD}. The coefficient which  characterizes the variation of 
 $\mathrm{\sqrt{\sigma_s}}$ with $\mathrm{g_E^2}$ is in perfect agreement with our 
 observation in $2+1$ flavor QCD. This confirms the fact that indeed the dynamical fermions do not 
 affect such observables at $\mathrm{T \gtrsim 5T_c}$, since it is most sensitive to the magnetic gluons 
 which interact non-perturbatively even at asymptotically high temperatures.  Our result agrees
 with and improves upon the previous lattice study~\cite{Boyd:1996bx} in 4 dimensional SU(3) gauge theory which reported 
 $\mathrm{\sqrt{\sigma_s} =0.566(13)~g_E^2}$, where the tree level matching $\mathrm{g_E^2=g^2(T) T}$ was 
 used in the fit but no continuum extrapolation was carried out.

Within 3 dimensional SU(3) gauge theory characterized by a coupling $\mathrm{g_3}$, 
two independent lattice studies have reported the variation of $\mathrm{\sqrt{\sigma_s}}$ as 
$\mathrm{\sim 0.553(2)g_3^2}$~\cite{Teper:1998te} and $\mathrm{\sim 0.554(4)g_3^2}$~\cite{Karsch:1994af} 
whereas a Hamiltonian-based approach in~\cite{Karabali:1998yq} parametrized it  
as $\mathrm{\sim 0.564~ g_3^2}$. In the temperature range $\mathrm{5<T/T_c< 7.3}$ we 
find that for $\mathrm{N_f=0}$ case, the $\mathrm{\sqrt{\sigma_s}/g_3^2=0.557(7)}$ has no noticeable 
dependence on temperature, which also alludes 
to the onset of dimensional reduction.  Here we have taken $\mathrm{g_3^2=g^2 T}$. We would similarly 
expect that the MQCD effective theory will start to describe the 2+1 flavor QCD data at these high 
temperatures as well.

\textbf{Consequences: Screening mass at high temperatures}
Having established the non-perturbative nature of the spatial Wilson line correlators
and the associated pseudo-potential and determined the value of the spatial string tension
the next question we ask is what are the corresponding consequences? In order to address this 
question one has to choose observables which are most sensitive to the magnetic scale. One of 
them is the spatial screening mass of hadron-like excitations~\cite{Detar:1987kae}. The screening 
correlators correspond to the long-wavelength or equivalently low-frequency modes of the high-temperature 
QCD plasma hence should be sensitive to the non-perturbative effects at the magnetic scale. The 
continuum extrapolated spatial screening mass of pseudo-scalar iso-triplet meson-like excitations 
in $2+1$ flavor QCD measured using HISQ discretization shows a significant deviation from the perturbative EQCD 
prediction~\cite{Laine:2003bd} at $\mathrm{T\lesssim 1}$ GeV~\cite{Bazavov:2019www,Brandt:2016daq,Cheng:2010fe} 
and more recently this deviation is observed to persist even at $\mathrm{T\lesssim 164}$ GeV~\cite{DallaBrida:2021ddx}. 
More interestingly there exist a clear difference in the magnitude of the vector and pseudo-scalar 
meson screening masses at such high temperatures~\cite{DallaBrida:2021ddx} which cannot be explained within 
the EQCD effective theory at $\mathrm{\mathcal{O}(g^2)}$.

\begin{figure}[H]
    \centering
    \includegraphics[width=0.8\textwidth]{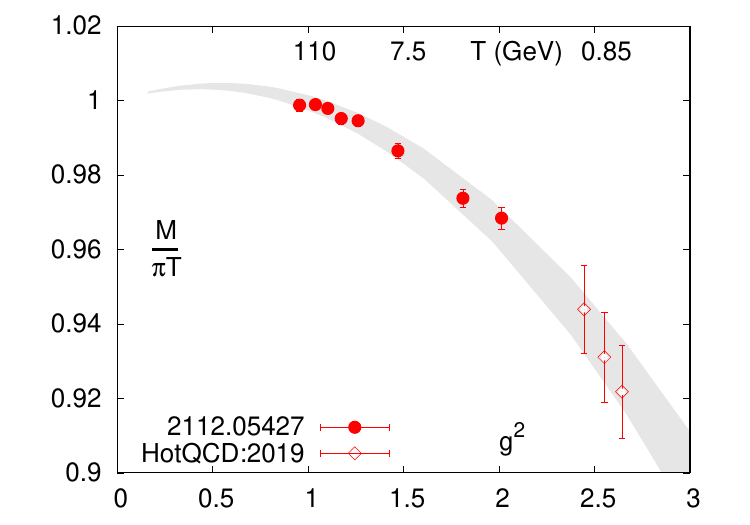}
    \caption{ The results for the mass $\mathrm{M}$ that goes into Eq.~\ref{eqn:SchrodingerEq}, obtained by performing 
    a fit to the spin-averaged data for screening mass from Ref.~\cite{DallaBrida:2021ddx} at $\mathrm{T\in [1,164]}$ GeV (filled points). The fit extrapolated to lower values of temperature provides a good match to the lattice data (open 
    points) from Ref.~\cite{Bazavov:2019www}.}
    \label{fig:VariationQuarkMassNP}
\end{figure}

In an attempt to understand these puzzles, we recall that at high enough temperatures quark excitations acquire 
an effective mass  $\mathrm{M=\pi T+g^2 T C_F/8\pi+\mathcal{O}(g^4)}$, and the spatial correlation function 
represent non-relativistic bound states of these heavy quarks in $2+1$ dimensions. The screening mass can thus be 
determined from the lowest eigenvalue of a two dimensional Schr\"{o}dinger equation with a static 
potential which until now, has been derived from a systematic \(\mathrm{1/(\pi T)}\) expansion of the heavy quark 
propagator in presence of magnetic gluons by resumming gluon-exchange ladder diagrams within EQCD~\cite{Laine:2003bd}. 
In order to include the effects of non-perturbatively interacting magnetic gluons
we propose a more general procedure to determine the potential by matching to the perturbative potential 
in the region $\mathrm{r < m_D^{-1}}$ to the continuum extrapolated potential $\mathrm{ V(r)}$ extracted in 
the previous section at \(\mathrm{r \sim m_D^{-1}}\).  Further details of this matching procedure can be found in 
supplementary material. The non-perturbative interactions also affect the effective mass 
$\mathrm{M}$. Hence we determine them by fitting the screening mass $\mathrm{2M+\delta m_{\text{scr}}}$ 
obtained by solving the Schr\"{o}dinger equation for the zero angular momentum state,
\begin{equation}\label{eqn:SchrodingerEq}
    \mathrm{\left[-\frac{1}{M}\left(\frac{d^2}{dr^2}+\frac{1}{r}\frac{d}{dr}\right)+V(r)\right]\psi_0(r)~=~\delta m_{\text{scr}}\psi_0(r)}~
\end{equation} 
for different trial values of $\mathrm{M}$, to the lattice data 
on the spin-averaged values of the 
continuum extrapolated screening mass in $2+1$ flavor QCD~\cite{DallaBrida:2021ddx} in the temperature 
range $\mathrm{T\in [1,164]}$ GeV. This is necessary due to the fact that the pseudo-scalar and vector 
screening masses are not degenerate, and split evenly due to spin-spin interaction~\cite{Eichten:1980mw,Koch:1992nx} 
between a heavy quark-antiquark pair which we will derive within the NRQCD framework starting from EQCD Lagrangian. 
The results for $\mathrm{M/(\pi T)}$ as a function of $\mathrm{g^2}$ is shown in Fig.~\ref{fig:VariationQuarkMassNP} which 
we could parametrize using $\mathrm{M/(\pi T)=1+0.0159(15) g^2-0.0165(10) g^4}$ with a $\mathrm{\chi^2/\text{d.o.f}= 
0.53}$. The coupling $\mathrm{g}$ was estimated at the scale $\mathrm{2\pi T}$ using the five-loop beta function with 
$\mathrm{\Lambda^{(3)}_{\overline{MS}}=0.339(12)}$. The coefficient of the $\mathrm{\mathcal{O}(g^2)}$ term is consistent 
with the perturbative estimate ($0.01688$) ~\cite{Laine:2003bd} but non-perturbative interactions which contribute at $
\mathrm{\mathcal{O}(g^4)}$ is also of the same magnitude with opposite sign. At asymptotically high temperatures $
\mathrm{M}$ goes over to $\mathrm{\pi T}$ as expected.

\begin{figure}[H]
    \centering
    \includegraphics[width=0.8\textwidth]{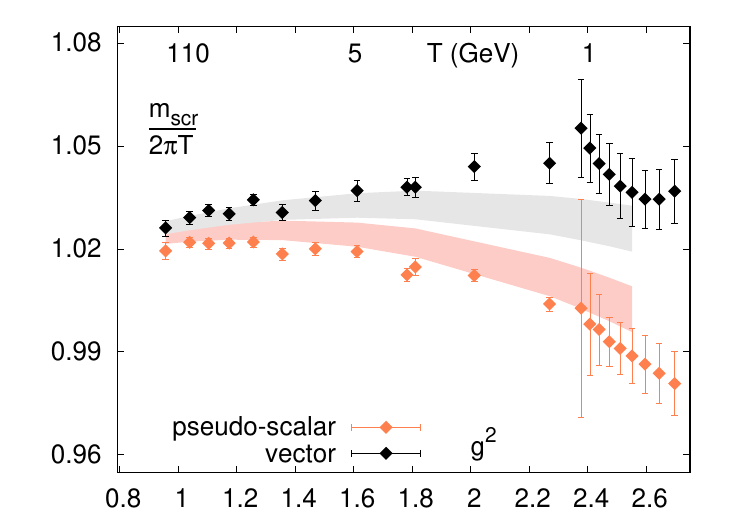}
    \caption{ The splitting between the pseudo-scalar and vector screening masses, shown as red and 
    gray bands respectively, predicted using the spin-spin interaction term derived perturbatively. Our results 
    are compared with lattice QCD data from Refs.~\cite{Bazavov:2019www,Giusti:2001xf}. The bands were drawn by performing 
    a simple interpolation of the data points in the entire range of temperatures $\mathrm{\gtrsim 5~T_c}$. }  
    \label{fig:PseudoscalarvectorscreeningmassesforHOTQCDdata}
\end{figure}

To check the robustness of our procedure we have used this parametrization to estimate spin-averaged screening masses 
temperatures lower than $1$ GeV, which are in good agreement with available lattice QCD results~\cite{Bazavov:2019www} 
upto $\mathrm{T \gtrsim 5T_c}$, where we earlier observed the onset of dimensional reduction.  We have also derived 
the spin dependent potential, $\mathrm{V_s(r)=\left(\pm\frac{1}{4}\right)\frac{4 g^2 T}{3M^2}\delta^2(r)}$ for vector and 
pseudo-scalar states respectively at $\mathrm{\mathcal{O}(g^2)}$, details of which are given in supplementary material.
Since this term is subdominant compared to our matched potential $\mathrm{V(r)}$, one can treat it as a perturbation.
This leads to a correction to the ground state energy of Eq.~\ref{eqn:SchrodingerEq} given by an amount 
$\mathrm{\delta m_{\text{scr}}^{(1)}(T)=\left(\pm\frac{1}{4}\right)\frac{2}{3\pi}\frac{g^2 T}{M^2}|\psi_0(0)|^2}$, 
for vector and pseudo-scalar states respectively, and depends on the ground state wave function $\mathrm{\psi_0}$ 
at the origin. We could then estimate the corresponding screening mass 
$\mathrm{2M+\delta m_{\text{scr}}(T)+\delta m_{\text{scr}}^{(1)}(T)}$ normalized by $\mathrm{2\pi T}$, which is 
shown in Fig.~\ref{fig:PseudoscalarvectorscreeningmassesforHOTQCDdata}.
Our calculated values are in excellent agreement with the lattice QCD 
data~\cite{Bazavov:2019www,Giusti:2001xf}, within $\mathrm{1\sigma}$ at 
$\mathrm{T\gtrsim 5T_c}$, but more importantly the correct $\mathrm{g^2}$ 
dependence is achieved over a large temperature range which could not be 
explained earlier. At lower temperatures, where dimensional reduction is not 
valid, we do not expect this spin potential to accurately describe 
the data.

\textbf{Summary \& Outlook:}
In this Letter, we explain an important feature of the strongly interacting quark gluon plasma i.e. the deviation of the 
screening mass corresponding to the long-wavelength excitations of the plasma from its perturbative EQCD prediction by 
including non-perturbative interactions of the magnetic gluons quantified by the spatial string tension. By performing 
a careful continuum extrapolation of $\mathrm{\sqrt{\sigma_s}}$ and studying its temperature dependence we estimate the 
temperature regime $\mathrm{T\gtrsim 5T_c}$ where QCD can be described by a dimensionally reduced effective theory.
In this regime we calculate the \emph{effective} potential that enters into the determination of the screening masses by 
matching the non-perturbative potential (at long distances) with the EQCD potential (valid at short distances) at the 
Debye scale. Remarkably the ground state wave function obtained using this potential can be used to explain the 
spin-splitting that exist between the pseudo-scalar and vector screening masses for a wide range of temperatures up to $
\mathrm{T\gtrsim 5T_c}$, without any additional parameter tuning. As a followup we will like to extend our formalism to 
explain the large spin-splitting that persist in the screening masses in QCD closer to $\mathrm{T_c}$ and further quantify 
how non-perturbative interactions at the magnetic scale enter into transport properties of the quark-gluon plasma. 
Another interesting direction will be to generalize our formalism to predict baryon screening masses~\cite{Giusti:2024ohu}.

\textbf{Acknowledgements}
This work was supported by the Deutsche Forschungsgemeinschaft
(DFG, German Research Foundation) Proj. No. 315477589-TRR 211 and 
the U.S. Department of Energy, Office of Science, through Contract No. DE-SC0012704. 
Computations have been performed on the GPU Cluster at Bielefeld University and 
at the Institute of Mathematical Sciences. Part of the HISQ calculations 
were performed using the \texttt{SIMULATeQCD}~\cite{HotQCD:2023ghu} 
library.

\section*{Appendix}

\subsection{Details of the Lattice methodology}
The spatial Wilson line correlators in $2+1$ flavor QCD at finite temperatures, are 
measured on the gauge configurations which were generated by HotQCD collaboration
to study QCD equation of state and quark number susceptibilities \cite{Bazavov:2011nk,HotQCD:2014kol,Bazavov:2013uja,Bazavov:2017dsy}
using physical strange quark mass and light quark mass $\mathrm{m_l = m_s/20}$ such 
that the Goldstone pion mass is $160$ MeV. These gauge configurations were generated using RHMC algorithm 
with highly improved staggered quarks action (HISQ) for the fermions~\cite{PhysRevD.75.054502,PhysRevD.59.074502}
and Symanzik improved gauge action. The correlators are calculated on configurations which are separated 
by $10$ RHMC steps. The choice of HISQ discretization ensures that finite cut-off and taste-splitting 
effects~\cite{PhysRevD.82.074501} are optimized even on a finite lattice. The lattice used in this study has a 
spatial volume $\mathrm{N^3}$ and number of sites along the Euclidean time direction to be $\mathrm{N_\tau}$ satisfying 
$\mathrm{N/N_\tau = 4}$ which ensures that finite volume effects are minimal. The temperature is denoted as 
$\mathrm{T=1/(N_\tau a)}$ hence at each temperature we have performed our measurements on $\mathrm{N_\tau = 8, 10}$ 
lattice in order to perform a continuum extrapolation of the spatial string tension extracted out of the spatial Wilson 
line correlators. The measurements are performed over a wide range of temperatures from $0.166 (0.168)$-$1(0.924)$ GeV 
for $\mathrm{N_{\tau} = 8~(10)}$ respectively and the typical lattice spacing ranges from $0.15$ fm to $0.04$ fm. 
The complete list of parameters used in our work are listed in Table~\ref{tab:dataqcd}. We have also performed 
the calculation of the spatial string tension on quenched SU(3) gauge configurations which are generated using the standard Wilson gauge action with heat-bath updates and 4 over-relaxation steps per update. The typical lattice extents along the spatial direction are chosen to $\mathrm{N = 32, 48, 64}$ and the corresponding $\mathrm{N_\tau = 8,12 , 16}$ for different $\mathrm{\beta = 6/g_0^2}$ values corresponding to the bare gauge coupling $\mathrm{g_0}$. The temperature 
corresponding to each lattice spacing has been calculated in terms of the Sommer parameter $\mathrm{r_0/a}$ whose parametrization in terms of $\beta$ has been taken from Ref.~\cite{Francis:2015lha}. 

\begin{table}
 \centering
 \begin{tabular}{|c|c|l|c|c|l|c|}
 \hline 
 $\beta$ & $\mathrm{N_\tau=8}$&   & &$\mathrm{N_\tau=10}$ &  &  \\
 \hline
 \hline
 & $\mathrm{T}$(MeV) &$\mathrm{N_{conf}}$& $\mathrm{\sigma_s/T^2}$  & $\mathrm{T}$(MeV) &$\mathrm{N_{conf}}$& $\mathrm{\sigma_s/T^2}$~~~~~ \\ 
  \hline
  6.423 & 166.6 &  1700 & 7.95(36) & & &\\
  \hline
6.488 &    177.5 &  1700 & 6.72(33) & & & \\
   \hline
6.515&    182.2 &  1700 & 6.58(11) & & &\\
   \hline
6.575 &  193.3 &  1700 & 5.75(26) & & & \\
  \hline
 6.664 & 210.7 &  1090 & 4.91(19) & 168.6  & 1071  & 7.75(14)\\ 
  \hline
6.800 &  240.3 &  1670 & 4.06(19) &  192.2  & 1835 & 5.59(26)\\
  \hline
6.950 &  277.2 &  2400 & 3.50(15) & 221.8 & 1317 & 4.50(20)\\
  \hline
7.150 &  334.2 &  850 & 2.90(4) & 267.4 & 1014 & 3.78(7)\\
  \hline
7.280 &  376.6 &  270 & 2.64(6)  & 301.3 & 828 & 3.23(9)\\
  \hline
7.373 &  409.7 &  500 & 2.50(9) & 327.8 &  869 & 3.05(9)\\
  \hline
7.596 &  500.0 &  500 & 2.03(5) &  400.0 & 672 &  2.58(6)\\
 \hline
7.825 &  611.1 &  500  & 1.79(5) &  488.9 &  717 & 2.17(6)\\
 \hline
8.000 &  711 &  1830  & 1.72(3) & & & \\
 \hline
8.200 &  843 &  249 & 1.47(6) &  675 &  900 &  1.82(3)\\
 \hline
 8.400 &  1000 & 249 & 1.41(5) &   800 & 750 & 1.63(3)\\
  \hline
8.570 & & & & 924 & 200 & 1.52(4)\\
 \hline
  \end{tabular}
  \caption{The parameters for the lattice calculations in $2+1$ flavor QCD for two different 
  $\mathrm{N_\tau=8,10}$ and the values of the spatial string tension $\mathrm{\sigma_s/T^2}$ at each temperature. }
  \label{tab:dataqcd}
\end{table}

\subsection{Details of the procedure to measure the Pseudo-potential and gauge-fixing}
To extract the color-singlet pseudo-potential between a infinitely massive, static $\mathrm{q\bar{q}}$ pair
separated by distance $\mathrm{r}$ we have defined the two-point spatial Wilson line correlator as,
\begin{equation}
\label{eqn:spatialwilsonlinecorrelator}
    \mathrm{W(r,l)}_{\vec{\mathrm{x}}_{\perp},\mathrm{z}}~\equiv~ \mathrm{\frac{1}{3}\textbf{Tr}_c}\left[\mathrm{S_z}(\vec{\mathrm{x}}_{\perp},\mathrm{l})~\mathrm{S_z}^\dagger(\vec{\mathrm{x}}_{\perp}+\vec{\mathrm{r}},\mathrm{l})\right]~,
\end{equation}
where $\mathrm{S_z}(\vec{\mathrm{x}}_{\perp},\tau,\mathrm{l})=\mathrm{\exp}\left({i\mathrm{\int_z^{z+l} dz~ A_z}(\vec{\mathrm{x}}_{\perp},\mathrm{z},\tau)}\right)$ 
is the spatial Wilson line along the $\mathrm{z}$ direction of length $\mathrm{l =L a}$ which are separated 
by a distance $\mathrm{r=R a}$ in the x-y plane at all possible initial $\mathrm{z}$ sites and on each $\tau$
slices. The $\mathrm{\textbf{Tr}_c}$ denotes trace over the color space. All possible such correlators 
consisting of Wilson lines of length $\mathrm{L}$ separated by $\mathrm{R}$ are measured and the average 
over these different sizes is denoted by $\mathrm{W(R,L)}$.  As the gauge noise present in 
the correlators defined at large $\mathrm{R,L}$ is typically large, we have only taken into account 
$\mathrm{R,L \leq N/2(N/4)}$ for on(off)-axis correlators.

The static potential between a static quark-antiquark pair has been calculated 
using temporal Wilson line correlators at zero temperature in a gauge-invariant 
manner in Ref.~\cite{Philipsen:2001ip,Philipsen:2002az}. In this 
calculation  one has to choose a gauge which is local in the Euclidean time 
such that the eigenspectrum of the transfer matrix remains exactly the same as
for the gauge invariant Wilson loop of the same size. Some of 
the possible gauge choices discussed are the Coulomb gauge and Laplacian Coulomb 
gauge. Typically, when determining the equation of state~\cite{PhysRevD.90.094503,PhysRevD.85.054503} 
the temperature is set in terms of a scale $\mathrm{r_0,r_1}$ which are extracted from 
the static potential in terms of temporal Wilson line correlators in the Coulomb 
gauge. In our case, we are calculating two-point spatial Wilson line correlator in 
the z direction, hence the choice of gauge should be such that it is local in the z 
direction to preserve the eigenstates of the spatial transfer matrix. We have chosen a 
modified-Coulomb gauge, which in the continuum limit is given by 
$\mathrm{\partial_x A_x +\partial_y A_y + \partial_\tau A_\tau = 0}$. 

\begin{figure}[H]
    \centering
    \includegraphics[width=0.6\textwidth]{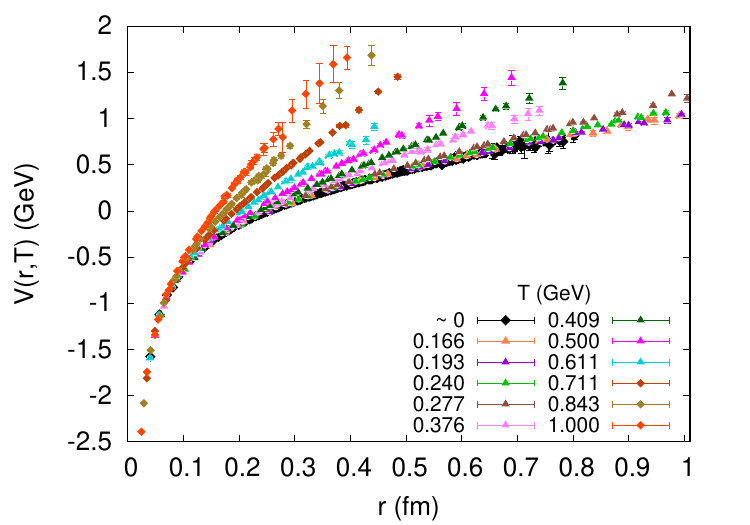}
    \includegraphics[width=0.6\textwidth]{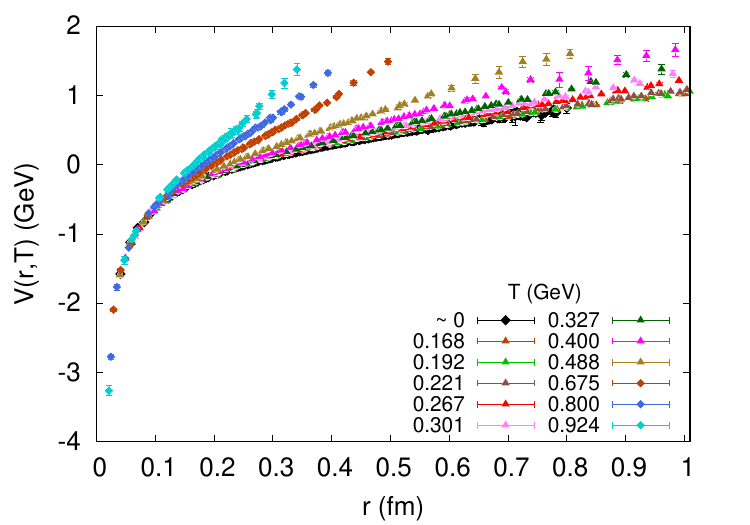}
    \caption{The pseudo-potential $\mathrm{V(r,T)}$ for $2+1$ flavor QCD as a function of $\mathrm{r}$ at two lattice 
    cut-offs corresponding to $\mathrm{N_{\tau} = 8, 10}$ is shown in the top and bottom panels respectively. The 
    temperature ranges from  $166$-$1000$ ($168-924$) MeV for $\mathrm{N_{\tau}=8~(10)}$. We have compared our 
    data with zero temperature potential denoted by black points, taken from Ref.~\cite{Brambilla:2022het}.}
   \label{fig:PaperPseudoPotentialNt8Nt10_2+1flavourQCD}
\end{figure}
We have implemented the gauge-fixing procedure as outlined in Ref.~\cite{Giusti:2001xf}. 
We extract the ground state energy out of the eigenvalues of the spatial Hamiltonian which 
in the limit $\mathrm{{L\rightarrow \infty}}$ gives us the pseudo-potential representing the 
interaction between a quark and the antiquark like excitations in addition to their self energies. 

The results for $\mathrm{V(r,T)}$ in physical units as a function of distance 
$\mathrm{r}$ is shown in Fig.~\ref{fig:PaperPseudoPotentialNt8Nt10_2+1flavourQCD} for different temperatures 
for $2+1$ flavor QCD and is compared to the zero temperature potential extracted earlier from 
Ref.~\cite{Brambilla:2022het}. It is clearly evident that the pseudo-potential do not vary with 
increasing temperatures up to about $\mathrm{T \approx 260}$ MeV, and this is 
independent of the lattice cut-off. Further the dependence of $\mathrm{V(r,T)}$ 
as a function of $\mathrm{r}$ is very similar to the zero temperature potential 
extracted from the temporal Wilson line correlators. Beyond $\mathrm{T \gtrsim 260}$ 
MeV the pseudo-potential rises more steeply as a function of $\mathrm{r}$ with increasing 
temperatures. Above $260$ MeV, the hard and the electric scales $\mathrm{\pi T, gT}$ just start 
to separate out, which is the reason behind the observable temperature dependence of the 
pseudo-potential. At the largest temperatures we have studied so-far the maximum 
extent of the box becomes small since the lattice spacing decreases. The finite 
volume effects at large distances are clearly visible resulting in a kink-like 
feature in $\mathrm{V(r,T)}$ at the highest temperatures, where we could reliably measure 
them. This is a characteristic feature arising from contamination of the ground state energy due to 
higher excited states. We have observed this while extracting the pseudo-potential by 
varying the $\mathrm{L_{min}}$ while pefoming a fit to the $\mathrm{m(L,R)}$. 
However large errors in the data at large $\mathrm{R}$ does not allow us to precisely 
determine the excited state contribution. We have not shown the pseudo-potentials 
for quenched QCD since the renormalization factor is not available in the literature.

\subsection{Details of the procedure to extract \texorpdfstring{$\mathrm{\alpha_L~\&~\sigma_s}$}{alpha=0} }

Within the dimensionally reduced effective theory of QCD, 
$\mathrm{\sqrt\sigma_s}$ increases linearly with temperature which 
also results in the increasing slope of the pseudo-potential with temperature
at large distances, evident in Fig.~\ref{fig:PaperPseudoPotentialNt8Nt10_2+1flavourQCD}. 
At first, we have extracted the coefficient of the $\mathrm{1/r}$ term using the ansatz 
$\mathrm{V(r) = V_0 + \sigma_s r - \alpha_c/r}$ over a range of values of 
$\mathrm{r\in [r_{min}:r_{max}]}$. Our strategy is to study the variation 
of $\mathrm{\alpha_c}$ as a function of $\mathrm{r_{min}}$. In this Cornell-like fit, the perturbative 
Coulomb contribution at $\mathrm{r<(\pi T)^{-1}}$ is expected to be sub-dominant at distances 
$\mathrm{r>(gT)^{-1}}$, where the L\"{u}scher term with strength $\mathrm{\alpha_L}$ 
will start to be important and will eventually be taken over by the string tension 
term at $\mathrm{r>(g^2 T/\pi)^{-1}}$. We have observed that the $\mathrm{\alpha_c}$ extracted 
from the fit decreases with increasing $\mathrm{r_\text{min}}$, where we have 
kept $\mathrm{r_\text{min}.T=1}$. The value of $\mathrm{\alpha_c}$ varies from its perturbative 
value $\mathrm{\alpha_P}$ at $\mathrm{r<(\pi T)^{-1}}$ to $\mathrm{\alpha_L}$ in the intermediate distances 
and eventually falling to zero at large distances $\mathrm{r\gtrsim (g^2T/\pi)^{-1}}$. We have thus 
extracted $\mathrm{\alpha_L}$ by varying our $\mathrm{r_{min}}$ in the range between 
$\mathrm{(\pi T)^{-1}-(g^2T/\pi)^{-1}}$ for both quenched QCD and $2+1$-flavor QCD at 
different temperatures. The values of the $\mathrm{\alpha_L}$ as a function of $\mathrm{T/T_c}$ 
are shown in Fig.~\ref{fig:Luscher term Pure gauge and 2+1 flavor QCD}. However extraction of 
$\mathrm{\alpha_L}$ with our procedure is not possible for temperatures $\mathrm{T\lesssim 2~T_c}$
as the scales are not well separated enough. 

Even though the coefficient of the $\mathrm{1/r}$ term goes to zero at large 
distances the extraction of $\mathrm{\sigma_s}$ from a 3 parameter 
Cornell-like fit is not very stable. Hence we set $\mathrm{r_{min}>(gT)^{-1}}$
and fixed the co-efficient of the $\mathrm{1/r}$ term to $\mathrm{i)~\alpha=\pi/12}$ or 
$\mathrm{ii)~\alpha=\pi/24}$ according to the temperature where we are 
performing the fit and extract $\mathrm{\sigma_s}$ from a two parameter fit. 
To verify the robustness of our procedure we have checked that the extracted 
$\mathrm{\sigma_s}$ is stable with the variation of $\mathrm{r_\text{min}}$ 
used in the fit. For low temperatures $\mathrm{T<4~T_c(1.5~T_c)}$, the $\mathrm{\sigma_s}$ 
is extracted from the two parameter fit keeping $\mathrm{\alpha_L=\pi/12}$ agrees with the 
Cornell-like fit value within 1-$\mathrm{\sigma}$. At higher temperatures i.e. $\mathrm{T>5~T_c(4~T_c)}$, 
the $\mathrm{\sigma_s}$ is obtained from a two parameter fit keeping $\mathrm{\alpha_L=\pi/24}$ fixed and 
it saturates to a constant value which also agrees to that obtained from the Cornell-like fit 
within 1-$\sigma$. This observation points toward the onset of dimensionally reduced effective 
theory description of QCD, both with and without dynamical fermions. The saturation in the extracted values 
of $\mathrm{\sigma_s}$ to a stable value at large $\mathrm{r_\text{min}}$ is even more evident for 
quenched QCD than in $2+1$ flavor QCD, where our statistics are higher. We have not considered the logarithm 
in r perturbative potential at short distances while performing the fits to the high temperature data, which would 
arise in 3 dimensions, as we cannot distinguish it from the Coulomb potential within our statistics.

\subsection{Details of matching procedure of \texorpdfstring{$\mathrm{V(R)}$}{alpha=0} with the EQCD inspired potential}

At high enough temperatures where the scales in thermal QCD are well separated i.e. 
$\mathrm{\pi T \gg gT}$, such the Euclidean temporal extent is infinitesimally small 
the $\mathrm{A_0}$ fields become static and QCD can be described in terms of an 
effective theory consisting of dynamical gauge fields $\mathrm{A_i}(\vec{\mathrm{x}})~,\mathrm{i \in [1,3]}$ 
coupled to adjoint scalar fields $\mathrm{A_0}(\vec{\mathrm{x}})$ and is known as EQCD
~\cite{Appelquist:1981vg,Braaten:1995jr}. The EQCD action in Euclidean space,
\begin{eqnarray}\label{EQCD Lagrangian}
    \mathrm{S_{\text{EQCD}}}&=&\mathrm{\int d^3x \left[\frac{1}{4}\bar{F}^a_{ij}\bar{F}^a_{ij}+\text{Tr}([\bar{D}_i,\bar{A}_4][\bar{D}_i,\bar{A}_4])
    \right.}\nonumber\\
    &&~~\mathrm{\left. +m_D^2\text{Tr}(\bar{A}_4^2)+\lambda_E(\text{Tr}(\bar{A}_4^2))^2+...\right]}, 
  \end{eqnarray}
can be written in terms of the low energy coefficients, the Debye mass $\mathrm{m_D}$ and couplings 
$\mathrm{g_E, \lambda_E}$ which at two loop in $\mathrm{g}$~\cite{Laine:2005ai} are
\begin{eqnarray}
    \mathrm{g_E^2}~&=&~\mathrm{ g^2T \left[1+\frac{g^2}{(4\pi)^2}\alpha_{E7}+\frac{g^4}{(4\pi)^4}\gamma_{E1}+\mathcal{O}(g^6)\right]} 
    \label{eqn:EQCDgaugecoupling}~\\
    \mathrm{m_D^2}~&=&~\mathrm{g^2T^2\left[\frac{N_c}{3}+\frac{N_f}{6}+\frac{g^2}{(4\pi)^2}\alpha_{E6}+\mathcal{O}(g^4)\right]}~. 
    \label{eqn:mDDebyeMass} 
    \end{eqnarray}
In our work the number of flavors is $\mathrm{N_f=3}$ and $\mathrm{N_c=3}$.
Note that $\mathrm{D_i=\partial_i - igA_i}$, $\mathrm{F_{ij}=\frac{i}{g}[D_i,D_j]}$ and 
the bar denotes fields defined within EQCD. Clearly the EQCD is valid at length scales 
$\mathrm{r>(\pi T)^{-1}}$. Since we are interested in the physics at infrared momentum 
scales $\mathrm{\lesssim g^2T/\pi}$, the effects of the fermion and gauge fields with 
non-zero Matsubara frequencies can be integrated out to produce an effective theory of the 
zero modes. The adjoint scalar fields which acquire a mass $\mathrm{\mathcal{O}(gT)}$ at 
the tree level in the static limit, can be integrated out to an effective theory valid 
at $\mathrm{r>(g^2T/\pi)^{-1}}$ known as MQCD, which is described by a 3 dimensional pure 
gauge action without dynamical fermions, 
\begin{eqnarray}
   \mathrm{S_{\text{MQCD}}}&=&\mathrm{\int d^3x \left(\frac{1}{4}\bar{\bar{F}}^a_{ij}\bar{\bar{F}}^a_{ij}+...\right),~~i,j=1,2,3}~.
\end{eqnarray}
with a dimensionful coupling $\mathrm{g_3}$ which can be related to the QCD coupling $\mathrm{g}$ in 4 dimensions 
through the following 2-loop relation~\cite{Giovannangeli:2003ti}.
When calculating the coupling $\mathrm{g_E}$ in the dimensionally reduced theory as a function of $\mathrm{g}$ 
using the Eq.~\ref{eqn:EQCDgaugecoupling}, the $\mathrm{g}$ is calculated at a scale $\mathrm{\bar\mu}$ which 
in the $\mathrm{\overline{MS}}$ scheme is $\mathrm{\bar{\mu}^2=4\pi \mu^2 e^{-\gamma_E}}$, 
where $\mathrm{\gamma_E}$ is the Euler-gamma constant and $\mathrm{\alpha_{E6}, \alpha_{E7}, \gamma_{E1}}$ 
depends on $\mathrm{\bar{\mu}/T}$. We have used the optimal-scale selection criterion~\cite{Laine:2005ai} where 
the scale $\mathrm{\mathrm{\mu_{opt}}}$ is chosen such that the error due to the arbitrariness in the scale 
setting for the determination of $\mathrm{g_E^2}$ remains minimal. The optimal scale is $\mathrm{\mu_{opt}/T=9.082}$ 
and $6.742$, while calculating $\mathrm{g_E^2}$ from $\mathrm{g}$ with $\mathrm{\Lambda^{(3)}_{\overline{MS}}=0.339(12)}$ and $
\mathrm{\Lambda^{(0)}_{\overline{MS}} =0.261(15)}$~\cite{FlavourLatticeAveragingGroupFLAG:2021npn} respectively in QCD 
with and without three dynamical fermions. We have used five loop beta function to estimate 
$\mathrm{g}$ at each temperature using the RunDec software~\cite{Schmidt:2012az}. To include all possible sources 
of errors in the calculation of $\mathrm{g_E}$ we have varied $\mathrm{\bar{\mu}\lesssim(0.5,2)\cdot\mu_{opt}}$ and taken into account 
the systematic errors due to uncertainties in $\mathrm{\Lambda_{\overline{MS}}}$ and $\mathrm{T_c}$, and the 
statistical errors in our estimation of $\mathrm{\sqrt{\sigma}_s}$ through a bootstrap analysis. 

\begin{figure}[H]
    \centering
    \includegraphics[width=0.8\textwidth]{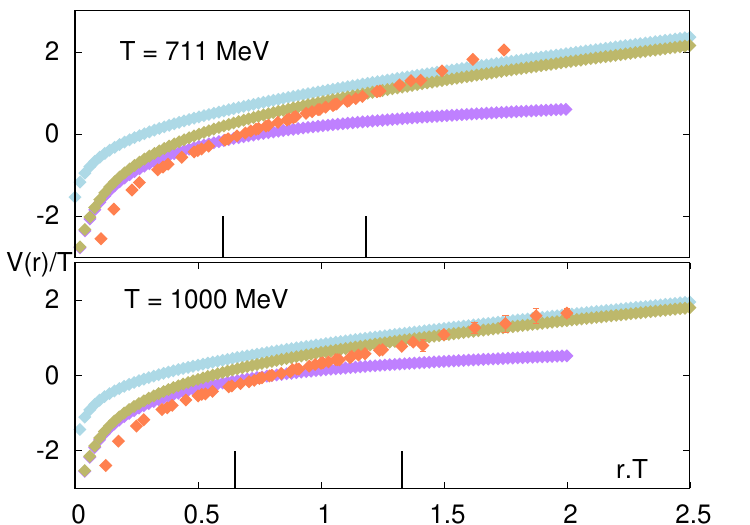}
    \caption{ Comparison of the non-perturbative pseudo-potential $\mathrm{V(r)}$ calculated in this work (orange) with
    the perturbative EQCD potential~\cite{Laine:2005ai}(purple) and the three dimensional perturbative potential derived 
    in Ref.~\cite{Schroder:1999sg} (light-blue). We have measured the complete potential by matching the EQCD potential 
    with our results for $\mathrm{V(r)}$ at the Debye scale given by $\mathrm{r\sim(gT)^{-1}}$ (left solid line). The second 
    black line to the right corresponds to non-perturbative magnetic scale $\mathrm{r=(g^2T/\pi)^{-1}}$.}
    \label{fig:Nonpert Pert potential comparison}
\end{figure}

Since the EQCD coupling is known from Eq.~\ref{eqn:mDDebyeMass}, we can perform a comparison of the spatial 
string tension calculated by us with the effective theory estimates, which has been discussed at length in the Letter. 
We now detail the procedure where we determine the most general pseudo-potential at high temperatures, by 
properly including the perturbative potential derived earlier within the EQCD framework~\cite{Laine:2005ai} 
and the non-perturbative part at the large distances encoded by the spatial string tension. Before proceeding 
we compare the perturbative EQCD potential with the pseudo-potential that we have estimated for 2+1 flavor QCD 
and also a variant of the potential calculated perturbatively in three dimensions~\cite{Schroder:1999sg}. The result 
for the comparison is shown in  Fig.~\ref{fig:Nonpert Pert potential comparison}. 
It is evident from the plot that the non-perturbative pseudo-potential (orange points)  differs from the EQCD potential 
(purple points) at short distances $\mathrm{r \lesssim 1/\pi T}$ but matches quite well in the intermediate region $\mathrm{(gT)^{-1} < r < (g^2T/\pi)^{-1}}$ and then starts again deviate from each other at large distances. This emphasizes 
the importance of non-perturbative effects at length scales larger than $\mathrm{(g^2 T/\pi)^{-1}}$. The perturbative 3 
dimensional potential start to agree with the non-perturbative V(r) at long distances only at very high temperatures 
$\gtrsim 1$ GeV. However since the ratio of the spatial string tension $\sigma_s$ measured in this work with/without 
dynamical fermions versus its perturbative estimate~\cite{Schroder:1999sg} is typically about $\mathrm{\sigma_s/
\sigma^{N_f=0}_{per}=2.30(2)}$ at $\mathrm{T\gtrsim 5T_c}$, the full non-perturbative determination of the pseudo-potential 
is essential to describe the physics of the deep magnetic sector of the quark-gluon plasma. Hence to correctly account for 
both the short distance perturbative interactions and the non-perturbative interactions at long distance scales we match 
the pseudo-potential $\mathrm{V(r)}$ measured in this work with the EQCD potential at $\mathrm{r\gtrsim1/m_D}$ which is 
implemented through the relations,
\begin{eqnarray}\label{EQCD potential}
    \mathrm{\frac{V(\hat{r})}{2\pi T}}~&=&~\mathrm{\frac{C_F}{(2\pi)^2}\frac{g_E^2}{T}\left[\ln{\frac{\hat{r}}{2}}+\gamma_E-K_0(\hat{r})\right],~~ \hat{r} < 1} \nonumber\\
    &=&~\mathrm{\frac{1}{2\pi}\left[\hat{\sigma_s} \frac{\hat{r}}{\hat{m}_D}~-~\frac{\pi}{24}\frac{\hat{m}_D}{\hat{r}}\right],~ \hat{r} \geq 1,}
\end{eqnarray}
where $\mathrm{\hat{r}\gtrsim rm_D}$, $\mathrm{C_F=\frac{N_c^2-1}{2N_c}}$, $\mathrm{\hat{\sigma}_s =\sigma_s/T^2}$, $\mathrm{\hat{m}_D=m_D/T}$, $\mathrm{\gamma_E}$ 
is Euler gamma-constant and $\mathrm{K_0(r)}$ is modified Bessel function of 2nd kind. This is done at temperatures 
$\mathrm{\gtrsim 5T_c}$ where the long distance part of the pseudo-potential is effectively described by MQCD.

\subsection{Details of Screening Mass calculation}
\label{sec:screeningmassdetails}
In the quark gluon plasma phase, color-singlet meson-like spatially correlated states describe the long 
wavelength plasma excitations~\cite{Detar:1987kae}. These \emph{screening} correlators can be constructed 
in different quantum number channels. One can predict the correlation lengths of these states or the corresponding 
screening masses $\mathrm{m_{scr}}$, by visualizing them as non-relativistic bound states of massive quark-like excitations 
of mass $\mathrm{\sim \pi T}$~\cite{Koch:1992nx} which are obtained by solving for the ground state of the 
Schr\"odinger equation,
\begin{equation}
\label{eqn:dimensionless Schrodinger eqn}
    \mathrm{\left[-\frac{\hat{m}^2_D}{2\pi^2 x}\left(\frac{d^2}{d\hat{r}^2}+\frac{1}{\hat{r}}\frac{d}{d\hat{r}}\right)+\frac{V(\hat{r})}{2\pi T}\right]\hat{\psi}_0(\hat{r})~=~\frac{\delta m_{scr}}{2\pi T}\hat{\psi}_0(\hat{r})}
\end{equation}
written in terms of a dimensionless variable $\mathrm{\hat r=m_D r}$ and a constant term $\mathrm{x=M/(\pi T)}$.
The ground state wave function is also normalized in units of $\mathrm{m_D}$,  
$\mathrm{\hat{\psi}_0(\hat{r})=\psi_0(\hat{r})/m_D}$  and the mass that enters in this calculation has 
to be appropriately renormalized. The renormalized mass has been calculated within 
EQCD~\cite{Laine:2003bd} as $\mathrm{M=\pi T+\frac{g^2 T}{8 \pi}C_F+\mathcal{O}(g^4T)}$.  The typical 
potential $\mathrm{V(r)}$ that enters into the calculations is usually derived perturbatively in EQCD where the 
parameters $\mathrm{g_E^2, m_D^2, M}$ are calculated upto $\mathcal{O}(g^2)$~\cite{Laine:2003bd}. 

In this work we have instead, used the potential calculated though our matching procedure discussed in 
the previous section in determining the screening mass. We also follow a different strategy, where we vary the 
input $\mathrm{M}$ while solving for Eq.~\ref{eqn:dimensionless Schrodinger eqn} such that the bound state energy 
$\mathrm{2M+\delta m_{scr}}$ thus obtained, agrees with the lattice QCD determination of spin-averaged meson 
screening masses in 2+1 flavor QCD at each temperature in the range from $\mathrm{T \approx 1}$-164 GeV~\cite{DallaBrida:2021ddx}. 
The spin-averaging has been done by averaging over the pseudo-scalar and vector screening mass since we will 
show in the next section that spin-dependent potential acts with equal magnitude but opposite signs between these 
states.
After extracting the $\mathrm{M}$ for these massive excitations, we have parametrized $\mathrm{M/(\pi T)}$ with an ansatz 
$\mathrm{1+c_1g^2+c_2g^4}$, where the coefficient of $\mathrm{g^4}$ term is a purely non-perturbative contribution, not studied 
earlier. Next using this parametrization of $\mathrm{M}$, we have solved the Schr\"{o}dinger equation for a new range of 
temperatures $\mathrm{T \leq 1}$ GeV and matched the ground state energy with the spin-averaged lattice QCD 
data for screening masses from Ref.~\cite{Bazavov:2019www}. We find an excellent agreement upto temperatures as low as $
\mathrm{5~T_c}$ ensuring the robustness of our procedure. Further we have explained the splitting that exist between the 
pseudo-scalar and vector screening masses which cannot be explained within the perturbative EQCD calculations. We will 
outline the derivation of the spin-spin interaction potential starting from a $2+1$-dimensional NRQCD Lagrangian, in the 
next section. 

\subsection{Derivation of spin dependent potential}
We start from the EQCD gauge fields whose Lagrangian in $2+1$ dimensions is given 
in Eq.~\ref{EQCD Lagrangian}  and introduce heavy fermion-like excitations with an 
effective mass $\mathrm{M}$. Using non-relativistic approximation i.e. expanding in powers of 
$\mathrm{\mathcal{O}(1/M)}$~\cite{Laine:2003bd} one obtains the NRQCD Lagrangian that denotes 
the dynamics of a fermion-like excitations. The effective Lagrangian in terms of two 
component spinors $\mathrm{\chi(\phi)}$ is
\begin{eqnarray}
    \mathrm{\mathcal{L}^f_{E}}&=&i\chi^\dagger\left[\mathrm{M -gA_0+D_3-\frac{1}{2M}}\left(\mathrm{D_k^2-}i\mathrm{g[s_k,s_l]F_{kl}}\right) \right]\chi \nonumber\\
    &+& i\phi^\dagger\left[\mathrm{M -gA_0-D_3-\frac{1}{2M}}\left(\mathrm{D_k^2-}i\mathrm{g[s_k,s_l]F_{kl}}\right) \right]\phi \nonumber~.
\end{eqnarray}
In the EQCD Lagrangian\ref{EQCD Lagrangian} gauge fields have 
mass dimension half after rescaling with $\mathrm{T^{1/2}}$ 
whereas in the above Lagrangian it is unity. The spinor fields 
with spin $\mathrm{s_k\equiv\sigma_k/2}$, labeled by the index $\mathrm{k=1,2}$, 
where $\mathrm{\sigma_k}$ are the Pauli matrices, are separated by a distance $\mathrm{r_{\perp}=|}\vec{\mathrm{x}}_{1{\perp}}-\vec{\mathrm{x}}_{2{\perp}}|$ 
and evolves along the $\mathrm{z}$ direction from $\mathrm{z=-\frac{L}{2}}$ to $\mathrm{z=\frac{L}{2}}$. The quantum state of 
this system can be constructed out of the vacuum $\mathrm{|\Omega \rangle}$, which is
\begin{eqnarray}
\nonumber
\mathrm{|\psi(z)\rangle \equiv \chi^\dagger_\beta}(\vec{\mathrm{x}}_{1{\perp}},\mathrm{z}) \Gamma^{\beta\alpha} \mathrm{U^\dagger(z},\vec{\mathrm{x}}_{1{\perp}},\vec{\mathrm{x}}_{2{\perp}})\phi_\alpha(\vec{\mathrm{x}}_{2{\perp}},z)|\Omega \rangle
\end{eqnarray}
The $\mathrm{\Gamma}$ matrix determines the quantum number of the meson-like channel we want to consider. The static 
potential corresponding to the spin-spin interaction can be extracted from the leading order term of the 
amplitude $\mathrm{\langle\psi\left(\frac{L}{2}\right)|\psi\left(-\frac{L}{2}\right)\rangle}$ that follows an 
exponential fall-off at large $\mathrm{L}$. Using a perturbative expansion in orders of $\mathrm{g}$ we can calculate this 
amplitude in the path integral method recovering the EQCD potential in Eq.~\ref{EQCD potential}.

\begin{figure}
    \centering
    \includegraphics[width=0.7\textwidth]{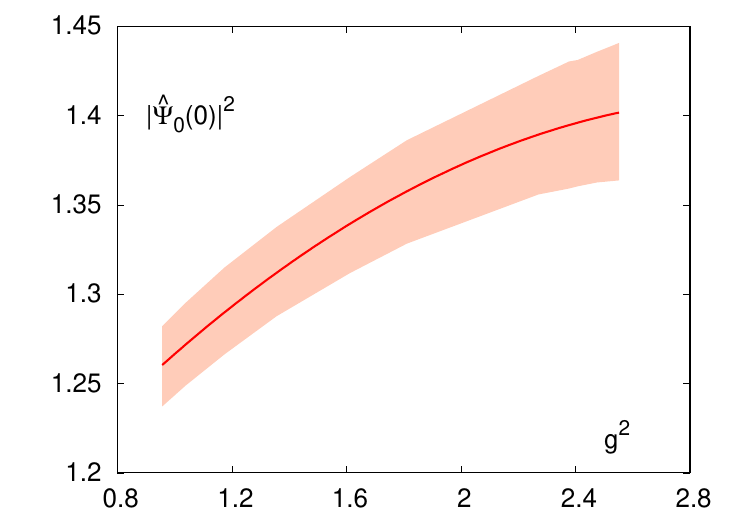}
    \caption{The probability density of the ground-state wave function representing a screening state at the origin can be parametrized 
    as $\mathrm{|\hat{\psi}_0(0)|^2=1.092(2)+0.208(3)g^2 -0.0340(9)g^4}$ within our non-perturbative treatment of the Schr\"{o}dinger equation given in Eq.~\ref{eqn:dimensionless Schrodinger eqn}.}
    \label{fig:dimensionless Wave function}
\end{figure}

The amplitude can be calculated to be
\begin{eqnarray}
\nonumber
    &&\mathrm{\langle\psi\left(\frac{L}{2}\right)|\psi\left(-\frac{L}{2}\right)\rangle
    \xrightarrow{L\to\infty}~2\text{exp}\left[-2ML\right.}\\
    &+&\mathrm{\left.V(r_\perp)L 
    -\frac{g^2T~C_F~L}{2M^2} \text{Tr}\left[\Gamma(s_1)_z\Gamma(s_2)_z\right] \delta^2(r_\perp)\right]}
\end{eqnarray}
For the pseudo-scalar ($\mathrm{\Gamma\equiv\sigma_3}$) and vector channels ($\mathrm{\Gamma\equiv\sigma_{1,2}}$) the  
potential turns out to be $\mathrm{V(r_{\perp})\mp\frac{g^2TC_F}{M^2} \frac{1}{4} \delta^2(r_{\perp})}
$ respectively, denoting the contribution due to spin-spin interaction.

A similar form of the spin-splitting potential was discussed earlier with different co-efficients for pseudo-scalar and 
vector channel in ref.~\cite{Koch:1992nx} but our formalism is more general. Since this term $\mathrm{V_s}$ is subdominant 
compared to $\mathrm{V(r)}$ already used for solving the Schr\"{o}dinger equation, we can treat it as a perturbation. We have 
also measured the strength of the bound state wave function (normalized by the Debye mass), at the origin 
$\mathrm{|\hat{\psi}_0(0)|^2}$ for different temperatures or equivalently the gauge couplings $\mathrm{g}$ 
shown in Fig.~\ref{fig:dimensionless Wave function}. We observe that the strength 
increases at higher couplings or equivalently lower temperatures, thus increasing 
the spin-splitting i.e. the difference between pseudo-scalar and vector screening 
masses, which is also evident in the lattice QCD data.

\bibliography{references_paper.bib}

\begin{thebibliography}{65}%
\makeatletter
\providecommand \@ifxundefined [1]{%
 \@ifx{#1\undefined}
}%
\providecommand \@ifnum [1]{%
 \ifnum #1\expandafter \@firstoftwo
 \else \expandafter \@secondoftwo
 \fi
}%
\providecommand \@ifx [1]{%
 \ifx #1\expandafter \@firstoftwo
 \else \expandafter \@secondoftwo
 \fi
}%
\providecommand \natexlab [1]{#1}%
\providecommand \enquote  [1]{``#1''}%
\providecommand \bibnamefont  [1]{#1}%
\providecommand \bibfnamefont [1]{#1}%
\providecommand \citenamefont [1]{#1}%
\providecommand \href@noop [0]{\@secondoftwo}%
\providecommand \href [0]{\begingroup \@sanitize@url \@href}%
\providecommand \@href[1]{\@@startlink{#1}\@@href}%
\providecommand \@@href[1]{\endgroup#1\@@endlink}%
\providecommand \@sanitize@url [0]{\catcode `\\12\catcode `\$12\catcode
  `\&12\catcode `\#12\catcode `\^12\catcode `\_12\catcode `\%12\relax}%
\providecommand \@@startlink[1]{}%
\providecommand \@@endlink[0]{}%
\providecommand \url  [0]{\begingroup\@sanitize@url \@url }%
\providecommand \@url [1]{\endgroup\@href {#1}{\urlprefix }}%
\providecommand \urlprefix  [0]{URL }%
\providecommand \Eprint [0]{\href }%
\providecommand \doibase [0]{http://dx.doi.org/}%
\providecommand \selectlanguage [0]{\@gobble}%
\providecommand \bibinfo  [0]{\@secondoftwo}%
\providecommand \bibfield  [0]{\@secondoftwo}%
\providecommand \translation [1]{[#1]}%
\providecommand \BibitemOpen [0]{}%
\providecommand \bibitemStop [0]{}%
\providecommand \bibitemNoStop [0]{.\EOS\space}%
\providecommand \EOS [0]{\spacefactor3000\relax}%
\providecommand \BibitemShut  [1]{\csname bibitem#1\endcsname}%
\let\auto@bib@innerbib\@empty
\bibitem [{\citenamefont {Gross}\ \emph {et~al.}(1981)\citenamefont {Gross},
  \citenamefont {Pisarski},\ and\ \citenamefont {Yaffe}}]{Gross:1980br}%
  \BibitemOpen
  \bibfield  {author} {\bibinfo {author} {\bibfnamefont {D.~J.}\ \bibnamefont
  {Gross}}, \bibinfo {author} {\bibfnamefont {R.~D.}\ \bibnamefont {Pisarski}},
  \ and\ \bibinfo {author} {\bibfnamefont {L.~G.}\ \bibnamefont {Yaffe}},\
  }\href {\doibase 10.1103/RevModPhys.53.43} {\bibfield  {journal} {\bibinfo
  {journal} {Rev. Mod. Phys.}\ }\textbf {\bibinfo {volume} {53}},\ \bibinfo
  {pages} {43} (\bibinfo {year} {1981})}\BibitemShut {NoStop}%
\bibitem [{\citenamefont {Shuryak}(1993)}]{Shuryak:1993kg}%
  \BibitemOpen
  \bibfield  {author} {\bibinfo {author} {\bibfnamefont {E.~V.}\ \bibnamefont
  {Shuryak}},\ }\href {\doibase 10.1103/RevModPhys.65.1} {\bibfield  {journal}
  {\bibinfo  {journal} {Rev. Mod. Phys.}\ }\textbf {\bibinfo {volume} {65}},\
  \bibinfo {pages} {1} (\bibinfo {year} {1993})}\BibitemShut {NoStop}%
\bibitem [{\citenamefont {Boyd}\ \emph {et~al.}(1995)\citenamefont {Boyd},
  \citenamefont {Engels}, \citenamefont {Karsch}, \citenamefont {Laermann},
  \citenamefont {Legeland}, \citenamefont {Lutgemeier},\ and\ \citenamefont
  {Petersson}}]{Boyd:1995zg}%
  \BibitemOpen
  \bibfield  {author} {\bibinfo {author} {\bibfnamefont {G.}~\bibnamefont
  {Boyd}}, \bibinfo {author} {\bibfnamefont {J.}~\bibnamefont {Engels}},
  \bibinfo {author} {\bibfnamefont {F.}~\bibnamefont {Karsch}}, \bibinfo
  {author} {\bibfnamefont {E.}~\bibnamefont {Laermann}}, \bibinfo {author}
  {\bibfnamefont {C.}~\bibnamefont {Legeland}}, \bibinfo {author}
  {\bibfnamefont {M.}~\bibnamefont {Lutgemeier}}, \ and\ \bibinfo {author}
  {\bibfnamefont {B.}~\bibnamefont {Petersson}},\ }\href {\doibase
  10.1103/PhysRevLett.75.4169} {\bibfield  {journal} {\bibinfo  {journal}
  {Phys. Rev. Lett.}\ }\textbf {\bibinfo {volume} {75}},\ \bibinfo {pages}
  {4169} (\bibinfo {year} {1995})},\ \Eprint
  {http://arxiv.org/abs/hep-lat/9506025} {arXiv:hep-lat/9506025} \BibitemShut
  {NoStop}%
\bibitem [{\citenamefont {Aoki}\ \emph {et~al.}(2006)\citenamefont {Aoki},
  \citenamefont {Endrodi}, \citenamefont {Fodor}, \citenamefont {Katz},\ and\
  \citenamefont {Szabo}}]{Aoki:2006we}%
  \BibitemOpen
  \bibfield  {author} {\bibinfo {author} {\bibfnamefont {Y.}~\bibnamefont
  {Aoki}}, \bibinfo {author} {\bibfnamefont {G.}~\bibnamefont {Endrodi}},
  \bibinfo {author} {\bibfnamefont {Z.}~\bibnamefont {Fodor}}, \bibinfo
  {author} {\bibfnamefont {S.~D.}\ \bibnamefont {Katz}}, \ and\ \bibinfo
  {author} {\bibfnamefont {K.~K.}\ \bibnamefont {Szabo}},\ }\href {\doibase
  10.1038/nature05120} {\bibfield  {journal} {\bibinfo  {journal} {Nature}\
  }\textbf {\bibinfo {volume} {443}},\ \bibinfo {pages} {675} (\bibinfo {year}
  {2006})},\ \Eprint {http://arxiv.org/abs/hep-lat/0611014}
  {arXiv:hep-lat/0611014} \BibitemShut {NoStop}%
\bibitem [{\citenamefont {Francis}\ \emph {et~al.}(2015)\citenamefont
  {Francis}, \citenamefont {Kaczmarek}, \citenamefont {Laine}, \citenamefont
  {Neuhaus},\ and\ \citenamefont {Ohno}}]{Francis:2015lha}%
  \BibitemOpen
  \bibfield  {author} {\bibinfo {author} {\bibfnamefont {A.}~\bibnamefont
  {Francis}}, \bibinfo {author} {\bibfnamefont {O.}~\bibnamefont {Kaczmarek}},
  \bibinfo {author} {\bibfnamefont {M.}~\bibnamefont {Laine}}, \bibinfo
  {author} {\bibfnamefont {T.}~\bibnamefont {Neuhaus}}, \ and\ \bibinfo
  {author} {\bibfnamefont {H.}~\bibnamefont {Ohno}},\ }\href {\doibase
  10.1103/PhysRevD.91.096002} {\bibfield  {journal} {\bibinfo  {journal} {Phys.
  Rev. D}\ }\textbf {\bibinfo {volume} {91}},\ \bibinfo {pages} {096002}
  (\bibinfo {year} {2015})},\ \Eprint {http://arxiv.org/abs/1503.05652}
  {arXiv:1503.05652 [hep-lat]} \BibitemShut {NoStop}%
\bibitem [{\citenamefont {Bazavov}\ \emph
  {et~al.}(2019{\natexlab{a}})\citenamefont {Bazavov}, \citenamefont {Ding},
  \citenamefont {Hegde}, \citenamefont {Kaczmarek}, \citenamefont {Karsch},
  \citenamefont {Karthik}, \citenamefont {Laermann}, \citenamefont {Lahiri},
  \citenamefont {Larsen}, \citenamefont {Li}, \citenamefont {Mukherjee},
  \citenamefont {Ohno}, \citenamefont {Petreczky}, \citenamefont {Sandmeyer},
  \citenamefont {Schmidt}, \citenamefont {Sharma},\ and\ \citenamefont
  {Steinbrecher}}]{HotQCD:2018pds}%
  \BibitemOpen
  \bibfield  {author} {\bibinfo {author} {\bibfnamefont {A.}~\bibnamefont
  {Bazavov}}, \bibinfo {author} {\bibfnamefont {H.-T.}\ \bibnamefont {Ding}},
  \bibinfo {author} {\bibfnamefont {P.}~\bibnamefont {Hegde}}, \bibinfo
  {author} {\bibfnamefont {O.}~\bibnamefont {Kaczmarek}}, \bibinfo {author}
  {\bibfnamefont {F.}~\bibnamefont {Karsch}}, \bibinfo {author} {\bibfnamefont
  {N.}~\bibnamefont {Karthik}}, \bibinfo {author} {\bibfnamefont
  {E.}~\bibnamefont {Laermann}}, \bibinfo {author} {\bibfnamefont
  {A.}~\bibnamefont {Lahiri}}, \bibinfo {author} {\bibfnamefont
  {R.}~\bibnamefont {Larsen}}, \bibinfo {author} {\bibfnamefont {S.-T.}\
  \bibnamefont {Li}}, \bibinfo {author} {\bibfnamefont {S.}~\bibnamefont
  {Mukherjee}}, \bibinfo {author} {\bibfnamefont {H.}~\bibnamefont {Ohno}},
  \bibinfo {author} {\bibfnamefont {P.}~\bibnamefont {Petreczky}}, \bibinfo
  {author} {\bibfnamefont {H.}~\bibnamefont {Sandmeyer}}, \bibinfo {author}
  {\bibfnamefont {C.}~\bibnamefont {Schmidt}}, \bibinfo {author} {\bibfnamefont
  {S.}~\bibnamefont {Sharma}}, \ and\ \bibinfo {author} {\bibfnamefont
  {P.}~\bibnamefont {Steinbrecher}},\ }\href {\doibase
  10.1016/j.physletb.2019.05.013} {\bibfield  {journal} {\bibinfo  {journal}
  {Physics Letters B}\ }\textbf {\bibinfo {volume} {795}},\ \bibinfo {pages}
  {15–21} (\bibinfo {year} {2019}{\natexlab{a}})}\BibitemShut {NoStop}%
\bibitem [{\citenamefont {Burger}\ \emph {et~al.}(2018)\citenamefont {Burger},
  \citenamefont {Ilgenfritz}, \citenamefont {Lombardo},\ and\ \citenamefont
  {Trunin}}]{Burger:2018fvb}%
  \BibitemOpen
  \bibfield  {author} {\bibinfo {author} {\bibfnamefont {F.}~\bibnamefont
  {Burger}}, \bibinfo {author} {\bibfnamefont {E.-M.}\ \bibnamefont
  {Ilgenfritz}}, \bibinfo {author} {\bibfnamefont {M.~P.}\ \bibnamefont
  {Lombardo}}, \ and\ \bibinfo {author} {\bibfnamefont {A.}~\bibnamefont
  {Trunin}},\ }\href {\doibase 10.1103/PhysRevD.98.094501} {\bibfield
  {journal} {\bibinfo  {journal} {Phys. Rev. D}\ }\textbf {\bibinfo {volume}
  {98}},\ \bibinfo {pages} {094501} (\bibinfo {year} {2018})},\ \Eprint
  {http://arxiv.org/abs/1805.06001} {arXiv:1805.06001 [hep-lat]} \BibitemShut
  {NoStop}%
\bibitem [{\citenamefont {Borsanyi}\ \emph {et~al.}(2020)\citenamefont
  {Borsanyi}, \citenamefont {Fodor}, \citenamefont {Guenther}, \citenamefont
  {Kara}, \citenamefont {Katz}, \citenamefont {Parotto}, \citenamefont
  {Pasztor}, \citenamefont {Ratti},\ and\ \citenamefont
  {Szabo}}]{Borsanyi:2020fev}%
  \BibitemOpen
  \bibfield  {author} {\bibinfo {author} {\bibfnamefont {S.}~\bibnamefont
  {Borsanyi}}, \bibinfo {author} {\bibfnamefont {Z.}~\bibnamefont {Fodor}},
  \bibinfo {author} {\bibfnamefont {J.~N.}\ \bibnamefont {Guenther}}, \bibinfo
  {author} {\bibfnamefont {R.}~\bibnamefont {Kara}}, \bibinfo {author}
  {\bibfnamefont {S.~D.}\ \bibnamefont {Katz}}, \bibinfo {author}
  {\bibfnamefont {P.}~\bibnamefont {Parotto}}, \bibinfo {author} {\bibfnamefont
  {A.}~\bibnamefont {Pasztor}}, \bibinfo {author} {\bibfnamefont
  {C.}~\bibnamefont {Ratti}}, \ and\ \bibinfo {author} {\bibfnamefont {K.~K.}\
  \bibnamefont {Szabo}},\ }\href {\doibase 10.1103/PhysRevLett.125.052001}
  {\bibfield  {journal} {\bibinfo  {journal} {Phys. Rev. Lett.}\ }\textbf
  {\bibinfo {volume} {125}},\ \bibinfo {pages} {052001} (\bibinfo {year}
  {2020})},\ \Eprint {http://arxiv.org/abs/2002.02821} {arXiv:2002.02821
  [hep-lat]} \BibitemShut {NoStop}%
\bibitem [{\citenamefont {Gavai}\ \emph {et~al.}(2024)\citenamefont {Gavai},
  \citenamefont {Jaensch}, \citenamefont {Kaczmarek}, \citenamefont {Karsch},
  \citenamefont {Sarkar}, \citenamefont {Shanker}, \citenamefont {Sharma},
  \citenamefont {Sharma},\ and\ \citenamefont {Ueding}}]{Gavai:2024mcj}%
  \BibitemOpen
  \bibfield  {author} {\bibinfo {author} {\bibfnamefont {R.~V.}\ \bibnamefont
  {Gavai}}, \bibinfo {author} {\bibfnamefont {M.~E.}\ \bibnamefont {Jaensch}},
  \bibinfo {author} {\bibfnamefont {O.}~\bibnamefont {Kaczmarek}}, \bibinfo
  {author} {\bibfnamefont {F.}~\bibnamefont {Karsch}}, \bibinfo {author}
  {\bibfnamefont {M.}~\bibnamefont {Sarkar}}, \bibinfo {author} {\bibfnamefont
  {R.}~\bibnamefont {Shanker}}, \bibinfo {author} {\bibfnamefont
  {S.}~\bibnamefont {Sharma}}, \bibinfo {author} {\bibfnamefont
  {S.}~\bibnamefont {Sharma}}, \ and\ \bibinfo {author} {\bibfnamefont
  {T.}~\bibnamefont {Ueding}},\ }\href@noop {} {\  (\bibinfo {year} {2024})},\
  \Eprint {http://arxiv.org/abs/2411.10217} {arXiv:2411.10217 [hep-lat]}
  \BibitemShut {NoStop}%
\bibitem [{\citenamefont {Appelquist}\ and\ \citenamefont
  {Carazzone}(1975)}]{Appelquist:1974tg}%
  \BibitemOpen
  \bibfield  {author} {\bibinfo {author} {\bibfnamefont {T.}~\bibnamefont
  {Appelquist}}\ and\ \bibinfo {author} {\bibfnamefont {J.}~\bibnamefont
  {Carazzone}},\ }\href {\doibase 10.1103/PhysRevD.11.2856} {\bibfield
  {journal} {\bibinfo  {journal} {Phys. Rev. D}\ }\textbf {\bibinfo {volume}
  {11}},\ \bibinfo {pages} {2856} (\bibinfo {year} {1975})}\BibitemShut
  {NoStop}%
\bibitem [{\citenamefont {Nadkarni}(1983)}]{Nadkarni:1982kb}%
  \BibitemOpen
  \bibfield  {author} {\bibinfo {author} {\bibfnamefont {S.}~\bibnamefont
  {Nadkarni}},\ }\href {\doibase 10.1103/PhysRevD.27.917} {\bibfield  {journal}
  {\bibinfo  {journal} {Phys. Rev. D}\ }\textbf {\bibinfo {volume} {27}},\
  \bibinfo {pages} {917} (\bibinfo {year} {1983})}\BibitemShut {NoStop}%
\bibitem [{\citenamefont {Braaten}\ and\ \citenamefont
  {Nieto}(1996)}]{Braaten:1995jr}%
  \BibitemOpen
  \bibfield  {author} {\bibinfo {author} {\bibfnamefont {E.}~\bibnamefont
  {Braaten}}\ and\ \bibinfo {author} {\bibfnamefont {A.}~\bibnamefont
  {Nieto}},\ }\href {\doibase 10.1103/PhysRevD.53.3421} {\bibfield  {journal}
  {\bibinfo  {journal} {Phys. Rev. D}\ }\textbf {\bibinfo {volume} {53}},\
  \bibinfo {pages} {3421} (\bibinfo {year} {1996})},\ \Eprint
  {http://arxiv.org/abs/hep-ph/9510408} {arXiv:hep-ph/9510408} \BibitemShut
  {NoStop}%
\bibitem [{\citenamefont {Kajantie}\ \emph {et~al.}(1997)\citenamefont
  {Kajantie}, \citenamefont {Laine}, \citenamefont {Rummukainen},\ and\
  \citenamefont {Shaposhnikov}}]{Kajantie:1997tt}%
  \BibitemOpen
  \bibfield  {author} {\bibinfo {author} {\bibfnamefont {K.}~\bibnamefont
  {Kajantie}}, \bibinfo {author} {\bibfnamefont {M.}~\bibnamefont {Laine}},
  \bibinfo {author} {\bibfnamefont {K.}~\bibnamefont {Rummukainen}}, \ and\
  \bibinfo {author} {\bibfnamefont {M.~E.}\ \bibnamefont {Shaposhnikov}},\
  }\href {\doibase 10.1016/S0550-3213(97)00425-2} {\bibfield  {journal}
  {\bibinfo  {journal} {Nucl. Phys. B}\ }\textbf {\bibinfo {volume} {503}},\
  \bibinfo {pages} {357} (\bibinfo {year} {1997})},\ \Eprint
  {http://arxiv.org/abs/hep-ph/9704416} {arXiv:hep-ph/9704416} \BibitemShut
  {NoStop}%
\bibitem [{\citenamefont {Karsch}\ \emph {et~al.}(1998)\citenamefont {Karsch},
  \citenamefont {Oevers},\ and\ \citenamefont {Petreczky}}]{Karsch:1998tx}%
  \BibitemOpen
  \bibfield  {author} {\bibinfo {author} {\bibfnamefont {F.}~\bibnamefont
  {Karsch}}, \bibinfo {author} {\bibfnamefont {M.}~\bibnamefont {Oevers}}, \
  and\ \bibinfo {author} {\bibfnamefont {P.}~\bibnamefont {Petreczky}},\ }\href
  {\doibase 10.1016/S0370-2693(98)01248-9} {\bibfield  {journal} {\bibinfo
  {journal} {Phys. Lett. B}\ }\textbf {\bibinfo {volume} {442}},\ \bibinfo
  {pages} {291} (\bibinfo {year} {1998})},\ \Eprint
  {http://arxiv.org/abs/hep-lat/9807035} {arXiv:hep-lat/9807035} \BibitemShut
  {NoStop}%
\bibitem [{\citenamefont {Linde}(1980)}]{Linde:1980ts}%
  \BibitemOpen
  \bibfield  {author} {\bibinfo {author} {\bibfnamefont {A.~D.}\ \bibnamefont
  {Linde}},\ }\href {\doibase 10.1016/0370-2693(80)90769-8} {\bibfield
  {journal} {\bibinfo  {journal} {Phys. Lett. B}\ }\textbf {\bibinfo {volume}
  {96}},\ \bibinfo {pages} {289} (\bibinfo {year} {1980})}\BibitemShut
  {NoStop}%
\bibitem [{\citenamefont {Appelquist}\ and\ \citenamefont
  {Pisarski}(1981)}]{Appelquist:1981vg}%
  \BibitemOpen
  \bibfield  {author} {\bibinfo {author} {\bibfnamefont {T.}~\bibnamefont
  {Appelquist}}\ and\ \bibinfo {author} {\bibfnamefont {R.~D.}\ \bibnamefont
  {Pisarski}},\ }\href {\doibase 10.1103/PhysRevD.23.2305} {\bibfield
  {journal} {\bibinfo  {journal} {Phys. Rev. D}\ }\textbf {\bibinfo {volume}
  {23}},\ \bibinfo {pages} {2305} (\bibinfo {year} {1981})}\BibitemShut
  {NoStop}%
\bibitem [{\citenamefont {D'Hoker}(1981)}]{DHoker:1980rnd}%
  \BibitemOpen
  \bibfield  {author} {\bibinfo {author} {\bibfnamefont {E.}~\bibnamefont
  {D'Hoker}},\ }\href {\doibase 10.1016/0550-3213(81)90425-9} {\bibfield
  {journal} {\bibinfo  {journal} {Nucl. Phys. B}\ }\textbf {\bibinfo {volume}
  {180}},\ \bibinfo {pages} {341} (\bibinfo {year} {1981})}\BibitemShut
  {NoStop}%
\bibitem [{\citenamefont {Laine}\ and\ \citenamefont
  {Philipsen}(1999)}]{Laine:1999hh}%
  \BibitemOpen
  \bibfield  {author} {\bibinfo {author} {\bibfnamefont {M.}~\bibnamefont
  {Laine}}\ and\ \bibinfo {author} {\bibfnamefont {O.}~\bibnamefont
  {Philipsen}},\ }\href {\doibase 10.1016/S0370-2693(99)00641-3} {\bibfield
  {journal} {\bibinfo  {journal} {Phys. Lett. B}\ }\textbf {\bibinfo {volume}
  {459}},\ \bibinfo {pages} {259} (\bibinfo {year} {1999})},\ \Eprint
  {http://arxiv.org/abs/hep-lat/9905004} {arXiv:hep-lat/9905004} \BibitemShut
  {NoStop}%
\bibitem [{\citenamefont {Gavai}\ and\ \citenamefont
  {Gupta}(2000)}]{Gavai:2000mx}%
  \BibitemOpen
  \bibfield  {author} {\bibinfo {author} {\bibfnamefont {R.~V.}\ \bibnamefont
  {Gavai}}\ and\ \bibinfo {author} {\bibfnamefont {S.}~\bibnamefont {Gupta}},\
  }\href {\doibase 10.1103/PhysRevLett.85.2068} {\bibfield  {journal} {\bibinfo
   {journal} {Phys. Rev. Lett.}\ }\textbf {\bibinfo {volume} {85}},\ \bibinfo
  {pages} {2068} (\bibinfo {year} {2000})},\ \Eprint
  {http://arxiv.org/abs/hep-lat/0004011} {arXiv:hep-lat/0004011} \BibitemShut
  {NoStop}%
\bibitem [{\citenamefont {Svetitsky}\ and\ \citenamefont
  {Yaffe}(1982)}]{Svetitsky:1982gs}%
  \BibitemOpen
  \bibfield  {author} {\bibinfo {author} {\bibfnamefont {B.}~\bibnamefont
  {Svetitsky}}\ and\ \bibinfo {author} {\bibfnamefont {L.~G.}\ \bibnamefont
  {Yaffe}},\ }\href {\doibase 10.1016/0550-3213(82)90172-9} {\bibfield
  {journal} {\bibinfo  {journal} {Nucl. Phys. B}\ }\textbf {\bibinfo {volume}
  {210}},\ \bibinfo {pages} {423} (\bibinfo {year} {1982})}\BibitemShut
  {NoStop}%
\bibitem [{\citenamefont {Borgs}(1985)}]{Borgs:1985qh}%
  \BibitemOpen
  \bibfield  {author} {\bibinfo {author} {\bibfnamefont {C.}~\bibnamefont
  {Borgs}},\ }\href {\doibase 10.1016/0550-3213(85)90582-6} {\bibfield
  {journal} {\bibinfo  {journal} {Nucl. Phys. B}\ }\textbf {\bibinfo {volume}
  {261}},\ \bibinfo {pages} {455} (\bibinfo {year} {1985})}\BibitemShut
  {NoStop}%
\bibitem [{\citenamefont {Manousakis}\ and\ \citenamefont
  {Polonyi}(1987)}]{Manousakis:1986jh}%
  \BibitemOpen
  \bibfield  {author} {\bibinfo {author} {\bibfnamefont {E.}~\bibnamefont
  {Manousakis}}\ and\ \bibinfo {author} {\bibfnamefont {J.}~\bibnamefont
  {Polonyi}},\ }\href {\doibase 10.1103/PhysRevLett.58.847} {\bibfield
  {journal} {\bibinfo  {journal} {Phys. Rev. Lett.}\ }\textbf {\bibinfo
  {volume} {58}},\ \bibinfo {pages} {847} (\bibinfo {year} {1987})}\BibitemShut
  {NoStop}%
\bibitem [{\citenamefont {Karkkainen}\ \emph {et~al.}(1993)\citenamefont
  {Karkkainen}, \citenamefont {Lacock}, \citenamefont {Miller}, \citenamefont
  {Petersson},\ and\ \citenamefont {Reisz}}]{Karkkainen:1993ch}%
  \BibitemOpen
  \bibfield  {author} {\bibinfo {author} {\bibfnamefont {L.}~\bibnamefont
  {Karkkainen}}, \bibinfo {author} {\bibfnamefont {P.}~\bibnamefont {Lacock}},
  \bibinfo {author} {\bibfnamefont {D.~E.}\ \bibnamefont {Miller}}, \bibinfo
  {author} {\bibfnamefont {B.}~\bibnamefont {Petersson}}, \ and\ \bibinfo
  {author} {\bibfnamefont {T.}~\bibnamefont {Reisz}},\ }\href {\doibase
  10.1016/0370-2693(93)90506-D} {\bibfield  {journal} {\bibinfo  {journal}
  {Phys. Lett. B}\ }\textbf {\bibinfo {volume} {312}},\ \bibinfo {pages} {173}
  (\bibinfo {year} {1993})},\ \Eprint {http://arxiv.org/abs/hep-lat/9306015}
  {arXiv:hep-lat/9306015} \BibitemShut {NoStop}%
\bibitem [{\citenamefont {Bali}\ \emph {et~al.}(1993)\citenamefont {Bali},
  \citenamefont {Fingberg}, \citenamefont {Heller}, \citenamefont {Karsch},\
  and\ \citenamefont {Schilling}}]{Bali:1993tz}%
  \BibitemOpen
  \bibfield  {author} {\bibinfo {author} {\bibfnamefont {G.~S.}\ \bibnamefont
  {Bali}}, \bibinfo {author} {\bibfnamefont {J.}~\bibnamefont {Fingberg}},
  \bibinfo {author} {\bibfnamefont {U.~M.}\ \bibnamefont {Heller}}, \bibinfo
  {author} {\bibfnamefont {F.}~\bibnamefont {Karsch}}, \ and\ \bibinfo {author}
  {\bibfnamefont {K.}~\bibnamefont {Schilling}},\ }\href {\doibase
  10.1103/PhysRevLett.71.3059} {\bibfield  {journal} {\bibinfo  {journal}
  {Phys. Rev. Lett.}\ }\textbf {\bibinfo {volume} {71}},\ \bibinfo {pages}
  {3059} (\bibinfo {year} {1993})},\ \Eprint
  {http://arxiv.org/abs/hep-lat/9306024} {arXiv:hep-lat/9306024} \BibitemShut
  {NoStop}%
\bibitem [{\citenamefont {Karsch}\ \emph {et~al.}(1995)\citenamefont {Karsch},
  \citenamefont {Laermann},\ and\ \citenamefont {Lütgemeier}}]{Karsch:1994af}%
  \BibitemOpen
  \bibfield  {author} {\bibinfo {author} {\bibfnamefont {F.}~\bibnamefont
  {Karsch}}, \bibinfo {author} {\bibfnamefont {E.}~\bibnamefont {Laermann}}, \
  and\ \bibinfo {author} {\bibfnamefont {M.}~\bibnamefont {Lütgemeier}},\
  }\href {\doibase 10.1016/0370-2693(94)01669-4} {\bibfield  {journal}
  {\bibinfo  {journal} {Physics Letters B}\ }\textbf {\bibinfo {volume}
  {346}},\ \bibinfo {pages} {94–98} (\bibinfo {year} {1995})}\BibitemShut
  {NoStop}%
\bibitem [{\citenamefont {Boyd}\ \emph {et~al.}(1996)\citenamefont {Boyd},
  \citenamefont {Engels}, \citenamefont {Karsch}, \citenamefont {Laermann},
  \citenamefont {Legeland}, \citenamefont {Lutgemeier},\ and\ \citenamefont
  {Petersson}}]{Boyd:1996bx}%
  \BibitemOpen
  \bibfield  {author} {\bibinfo {author} {\bibfnamefont {G.}~\bibnamefont
  {Boyd}}, \bibinfo {author} {\bibfnamefont {J.}~\bibnamefont {Engels}},
  \bibinfo {author} {\bibfnamefont {F.}~\bibnamefont {Karsch}}, \bibinfo
  {author} {\bibfnamefont {E.}~\bibnamefont {Laermann}}, \bibinfo {author}
  {\bibfnamefont {C.}~\bibnamefont {Legeland}}, \bibinfo {author}
  {\bibfnamefont {M.}~\bibnamefont {Lutgemeier}}, \ and\ \bibinfo {author}
  {\bibfnamefont {B.}~\bibnamefont {Petersson}},\ }\href {\doibase
  10.1016/0550-3213(96)00170-8} {\bibfield  {journal} {\bibinfo  {journal}
  {Nucl. Phys. B}\ }\textbf {\bibinfo {volume} {469}},\ \bibinfo {pages} {419}
  (\bibinfo {year} {1996})},\ \Eprint {http://arxiv.org/abs/hep-lat/9602007}
  {arXiv:hep-lat/9602007} \BibitemShut {NoStop}%
\bibitem [{\citenamefont {Cheng}\ \emph
  {et~al.}(2008{\natexlab{a}})\citenamefont {Cheng} \emph
  {et~al.}}]{Cheng:2008bs}%
  \BibitemOpen
  \bibfield  {author} {\bibinfo {author} {\bibfnamefont {M.}~\bibnamefont
  {Cheng}} \emph {et~al.},\ }\href {\doibase 10.1103/PhysRevD.78.034506}
  {\bibfield  {journal} {\bibinfo  {journal} {Phys. Rev. D}\ }\textbf {\bibinfo
  {volume} {78}},\ \bibinfo {pages} {034506} (\bibinfo {year}
  {2008}{\natexlab{a}})},\ \Eprint {http://arxiv.org/abs/0806.3264}
  {arXiv:0806.3264 [hep-lat]} \BibitemShut {NoStop}%
\bibitem [{\citenamefont {Haque}\ and\ \citenamefont
  {Mustafa}(2025)}]{Haque:2024gva}%
  \BibitemOpen
  \bibfield  {author} {\bibinfo {author} {\bibfnamefont {N.}~\bibnamefont
  {Haque}}\ and\ \bibinfo {author} {\bibfnamefont {M.~G.}\ \bibnamefont
  {Mustafa}},\ }\href {\doibase 10.1016/j.ppnp.2024.104136} {\bibfield
  {journal} {\bibinfo  {journal} {Prog. Part. Nucl. Phys.}\ }\textbf {\bibinfo
  {volume} {140}},\ \bibinfo {pages} {104136} (\bibinfo {year} {2025})},\
  \Eprint {http://arxiv.org/abs/2404.08734} {arXiv:2404.08734 [hep-ph]}
  \BibitemShut {NoStop}%
\bibitem [{\citenamefont {Laine}\ and\ \citenamefont
  {Vepsal1ainen}(2004)}]{Laine:2003bd}%
  \BibitemOpen
  \bibfield  {author} {\bibinfo {author} {\bibfnamefont {M.}~\bibnamefont
  {Laine}}\ and\ \bibinfo {author} {\bibfnamefont {M.}~\bibnamefont
  {Vepsal1ainen}},\ }\href {\doibase 10.1088/1126-6708/2004/02/004} {\bibfield
  {journal} {\bibinfo  {journal} {JHEP}\ }\textbf {\bibinfo {volume} {02}},\
  \bibinfo {pages} {004} (\bibinfo {year} {2004})},\ \Eprint
  {http://arxiv.org/abs/hep-ph/0311268} {arXiv:hep-ph/0311268} \BibitemShut
  {NoStop}%
\bibitem [{\citenamefont {Aubin}\ \emph {et~al.}(2004)\citenamefont {Aubin},
  \citenamefont {Bernard}, \citenamefont {DeTar}, \citenamefont {Osborn},
  \citenamefont {Gottlieb}, \citenamefont {Gregory}, \citenamefont {Toussaint},
  \citenamefont {Heller}, \citenamefont {Hetrick},\ and\ \citenamefont
  {Sugar}}]{Aubin:2004wf}%
  \BibitemOpen
  \bibfield  {author} {\bibinfo {author} {\bibfnamefont {C.}~\bibnamefont
  {Aubin}}, \bibinfo {author} {\bibfnamefont {C.}~\bibnamefont {Bernard}},
  \bibinfo {author} {\bibfnamefont {C.}~\bibnamefont {DeTar}}, \bibinfo
  {author} {\bibfnamefont {J.}~\bibnamefont {Osborn}}, \bibinfo {author}
  {\bibfnamefont {S.}~\bibnamefont {Gottlieb}}, \bibinfo {author}
  {\bibfnamefont {E.~B.}\ \bibnamefont {Gregory}}, \bibinfo {author}
  {\bibfnamefont {D.}~\bibnamefont {Toussaint}}, \bibinfo {author}
  {\bibfnamefont {U.~M.}\ \bibnamefont {Heller}}, \bibinfo {author}
  {\bibfnamefont {J.~E.}\ \bibnamefont {Hetrick}}, \ and\ \bibinfo {author}
  {\bibfnamefont {R.}~\bibnamefont {Sugar}},\ }\href {\doibase
  10.1103/PhysRevD.70.094505} {\bibfield  {journal} {\bibinfo  {journal} {Phys.
  Rev. D}\ }\textbf {\bibinfo {volume} {70}},\ \bibinfo {pages} {094505}
  (\bibinfo {year} {2004})},\ \Eprint {http://arxiv.org/abs/hep-lat/0402030}
  {arXiv:hep-lat/0402030} \BibitemShut {NoStop}%
\bibitem [{\citenamefont {Bazavov}\ \emph
  {et~al.}(2012{\natexlab{a}})\citenamefont {Bazavov} \emph
  {et~al.}}]{Bazavov:2011nk}%
  \BibitemOpen
  \bibfield  {author} {\bibinfo {author} {\bibfnamefont {A.}~\bibnamefont
  {Bazavov}} \emph {et~al.},\ }\href {\doibase 10.1103/PhysRevD.85.054503}
  {\bibfield  {journal} {\bibinfo  {journal} {Phys. Rev. D}\ }\textbf {\bibinfo
  {volume} {85}},\ \bibinfo {pages} {054503} (\bibinfo {year}
  {2012}{\natexlab{a}})},\ \Eprint {http://arxiv.org/abs/1111.1710}
  {arXiv:1111.1710 [hep-lat]} \BibitemShut {NoStop}%
\bibitem [{\citenamefont {Bazavov}\ \emph
  {et~al.}(2014{\natexlab{a}})\citenamefont {Bazavov} \emph
  {et~al.}}]{HotQCD:2014kol}%
  \BibitemOpen
  \bibfield  {author} {\bibinfo {author} {\bibfnamefont {A.}~\bibnamefont
  {Bazavov}} \emph {et~al.} (\bibinfo {collaboration} {HotQCD}),\ }\href
  {\doibase 10.1103/PhysRevD.90.094503} {\bibfield  {journal} {\bibinfo
  {journal} {Phys. Rev. D}\ }\textbf {\bibinfo {volume} {90}},\ \bibinfo
  {pages} {094503} (\bibinfo {year} {2014}{\natexlab{a}})},\ \Eprint
  {http://arxiv.org/abs/1407.6387} {arXiv:1407.6387 [hep-lat]} \BibitemShut
  {NoStop}%
\bibitem [{\citenamefont {Bazavov}\ \emph {et~al.}(2018)\citenamefont
  {Bazavov}, \citenamefont {Petreczky},\ and\ \citenamefont
  {Weber}}]{Bazavov:2017dsy}%
  \BibitemOpen
  \bibfield  {author} {\bibinfo {author} {\bibfnamefont {A.}~\bibnamefont
  {Bazavov}}, \bibinfo {author} {\bibfnamefont {P.}~\bibnamefont {Petreczky}},
  \ and\ \bibinfo {author} {\bibfnamefont {J.~H.}\ \bibnamefont {Weber}},\
  }\href {\doibase 10.1103/PhysRevD.97.014510} {\bibfield  {journal} {\bibinfo
  {journal} {Phys. Rev. D}\ }\textbf {\bibinfo {volume} {97}},\ \bibinfo
  {pages} {014510} (\bibinfo {year} {2018})},\ \Eprint
  {http://arxiv.org/abs/1710.05024} {arXiv:1710.05024 [hep-lat]} \BibitemShut
  {NoStop}%
\bibitem [{\citenamefont {Brambilla}\ \emph {et~al.}(2023)\citenamefont
  {Brambilla}, \citenamefont {Delgado}, \citenamefont {Kronfeld}, \citenamefont
  {Leino}, \citenamefont {Petreczky}, \citenamefont {Steinbei\ss{}er},
  \citenamefont {Vairo},\ and\ \citenamefont {Weber}}]{Brambilla:2022het}%
  \BibitemOpen
  \bibfield  {author} {\bibinfo {author} {\bibfnamefont {N.}~\bibnamefont
  {Brambilla}}, \bibinfo {author} {\bibfnamefont {R.~L.}\ \bibnamefont
  {Delgado}}, \bibinfo {author} {\bibfnamefont {A.~S.}\ \bibnamefont
  {Kronfeld}}, \bibinfo {author} {\bibfnamefont {V.}~\bibnamefont {Leino}},
  \bibinfo {author} {\bibfnamefont {P.}~\bibnamefont {Petreczky}}, \bibinfo
  {author} {\bibfnamefont {S.}~\bibnamefont {Steinbei\ss{}er}}, \bibinfo
  {author} {\bibfnamefont {A.}~\bibnamefont {Vairo}}, \ and\ \bibinfo {author}
  {\bibfnamefont {J.~H.}\ \bibnamefont {Weber}} (\bibinfo {collaboration}
  {TUMQCD}),\ }\href {\doibase 10.1103/PhysRevD.107.074503} {\bibfield
  {journal} {\bibinfo  {journal} {Phys. Rev. D}\ }\textbf {\bibinfo {volume}
  {107}},\ \bibinfo {pages} {074503} (\bibinfo {year} {2023})},\ \Eprint
  {http://arxiv.org/abs/2206.03156} {arXiv:2206.03156 [hep-lat]} \BibitemShut
  {NoStop}%
\bibitem [{\citenamefont {Cheng}\ \emph
  {et~al.}(2008{\natexlab{b}})\citenamefont {Cheng} \emph
  {et~al.}}]{Cheng:2007jq}%
  \BibitemOpen
  \bibfield  {author} {\bibinfo {author} {\bibfnamefont {M.}~\bibnamefont
  {Cheng}} \emph {et~al.},\ }\href {\doibase 10.1103/PhysRevD.77.014511}
  {\bibfield  {journal} {\bibinfo  {journal} {Phys. Rev. D}\ }\textbf {\bibinfo
  {volume} {77}},\ \bibinfo {pages} {014511} (\bibinfo {year}
  {2008}{\natexlab{b}})},\ \Eprint {http://arxiv.org/abs/0710.0354}
  {arXiv:0710.0354 [hep-lat]} \BibitemShut {NoStop}%
\bibitem [{\citenamefont {Bazavov}\ \emph {et~al.}(2016)\citenamefont
  {Bazavov}, \citenamefont {Brambilla}, \citenamefont {Ding}, \citenamefont
  {Petreczky}, \citenamefont {Schadler}, \citenamefont {Vairo},\ and\
  \citenamefont {Weber}}]{Bazavov:2016uvm}%
  \BibitemOpen
  \bibfield  {author} {\bibinfo {author} {\bibfnamefont {A.}~\bibnamefont
  {Bazavov}}, \bibinfo {author} {\bibfnamefont {N.}~\bibnamefont {Brambilla}},
  \bibinfo {author} {\bibfnamefont {H.~T.}\ \bibnamefont {Ding}}, \bibinfo
  {author} {\bibfnamefont {P.}~\bibnamefont {Petreczky}}, \bibinfo {author}
  {\bibfnamefont {H.~P.}\ \bibnamefont {Schadler}}, \bibinfo {author}
  {\bibfnamefont {A.}~\bibnamefont {Vairo}}, \ and\ \bibinfo {author}
  {\bibfnamefont {J.~H.}\ \bibnamefont {Weber}},\ }\href {\doibase
  10.1103/PhysRevD.93.114502} {\bibfield  {journal} {\bibinfo  {journal} {Phys.
  Rev. D}\ }\textbf {\bibinfo {volume} {93}},\ \bibinfo {pages} {114502}
  (\bibinfo {year} {2016})},\ \Eprint {http://arxiv.org/abs/1603.06637}
  {arXiv:1603.06637 [hep-lat]} \BibitemShut {NoStop}%
\bibitem [{\citenamefont {Luscher}(1981)}]{Luscher:1980ac}%
  \BibitemOpen
  \bibfield  {author} {\bibinfo {author} {\bibfnamefont {M.}~\bibnamefont
  {Luscher}},\ }\href {\doibase 10.1016/0550-3213(81)90423-5} {\bibfield
  {journal} {\bibinfo  {journal} {Nucl. Phys. B}\ }\textbf {\bibinfo {volume}
  {180}},\ \bibinfo {pages} {317} (\bibinfo {year} {1981})}\BibitemShut
  {NoStop}%
\bibitem [{\citenamefont {Luscher}\ and\ \citenamefont
  {Weisz}(2002)}]{Luscher:2002qv}%
  \BibitemOpen
  \bibfield  {author} {\bibinfo {author} {\bibfnamefont {M.}~\bibnamefont
  {Luscher}}\ and\ \bibinfo {author} {\bibfnamefont {P.}~\bibnamefont
  {Weisz}},\ }\href {\doibase 10.1088/1126-6708/2002/07/049} {\bibfield
  {journal} {\bibinfo  {journal} {JHEP}\ }\textbf {\bibinfo {volume} {07}},\
  \bibinfo {pages} {049} (\bibinfo {year} {2002})},\ \Eprint
  {http://arxiv.org/abs/hep-lat/0207003} {arXiv:hep-lat/0207003} \BibitemShut
  {NoStop}%
\bibitem [{\citenamefont {Luscher}\ \emph {et~al.}(1980)\citenamefont
  {Luscher}, \citenamefont {Symanzik},\ and\ \citenamefont
  {Weisz}}]{Luscher:1980fr}%
  \BibitemOpen
  \bibfield  {author} {\bibinfo {author} {\bibfnamefont {M.}~\bibnamefont
  {Luscher}}, \bibinfo {author} {\bibfnamefont {K.}~\bibnamefont {Symanzik}}, \
  and\ \bibinfo {author} {\bibfnamefont {P.}~\bibnamefont {Weisz}},\ }\href
  {\doibase 10.1016/0550-3213(80)90009-7} {\bibfield  {journal} {\bibinfo
  {journal} {Nucl. Phys. B}\ }\textbf {\bibinfo {volume} {173}},\ \bibinfo
  {pages} {365} (\bibinfo {year} {1980})}\BibitemShut {NoStop}%
\bibitem [{\citenamefont {Alvarez}(1981)}]{Alvarez:1981kc}%
  \BibitemOpen
  \bibfield  {author} {\bibinfo {author} {\bibfnamefont {O.}~\bibnamefont
  {Alvarez}},\ }\href {\doibase 10.1103/PhysRevD.24.440} {\bibfield  {journal}
  {\bibinfo  {journal} {Phys. Rev. D}\ }\textbf {\bibinfo {volume} {24}},\
  \bibinfo {pages} {440} (\bibinfo {year} {1981})}\BibitemShut {NoStop}%
\bibitem [{\citenamefont {Laine}\ and\ \citenamefont
  {Schroder}(2005)}]{Laine:2005ai}%
  \BibitemOpen
  \bibfield  {author} {\bibinfo {author} {\bibfnamefont {M.}~\bibnamefont
  {Laine}}\ and\ \bibinfo {author} {\bibfnamefont {Y.}~\bibnamefont
  {Schroder}},\ }\href {\doibase 10.1088/1126-6708/2005/03/067} {\bibfield
  {journal} {\bibinfo  {journal} {JHEP}\ }\textbf {\bibinfo {volume} {03}},\
  \bibinfo {pages} {067} (\bibinfo {year} {2005})},\ \Eprint
  {http://arxiv.org/abs/hep-ph/0503061} {arXiv:hep-ph/0503061} \BibitemShut
  {NoStop}%
\bibitem [{\citenamefont {Teper}(1999)}]{Teper:1998te}%
  \BibitemOpen
  \bibfield  {author} {\bibinfo {author} {\bibfnamefont {M.~J.}\ \bibnamefont
  {Teper}},\ }\href {\doibase 10.1103/PhysRevD.59.014512} {\bibfield  {journal}
  {\bibinfo  {journal} {Phys. Rev. D}\ }\textbf {\bibinfo {volume} {59}},\
  \bibinfo {pages} {014512} (\bibinfo {year} {1999})},\ \Eprint
  {http://arxiv.org/abs/hep-lat/9804008} {arXiv:hep-lat/9804008} \BibitemShut
  {NoStop}%
\bibitem [{\citenamefont {Karabali}\ \emph {et~al.}(1998)\citenamefont
  {Karabali}, \citenamefont {Kim},\ and\ \citenamefont
  {Nair}}]{Karabali:1998yq}%
  \BibitemOpen
  \bibfield  {author} {\bibinfo {author} {\bibfnamefont {D.}~\bibnamefont
  {Karabali}}, \bibinfo {author} {\bibfnamefont {C.-j.}\ \bibnamefont {Kim}}, \
  and\ \bibinfo {author} {\bibfnamefont {V.~P.}\ \bibnamefont {Nair}},\ }\href
  {\doibase 10.1016/S0370-2693(98)00751-5} {\bibfield  {journal} {\bibinfo
  {journal} {Phys. Lett. B}\ }\textbf {\bibinfo {volume} {434}},\ \bibinfo
  {pages} {103} (\bibinfo {year} {1998})},\ \Eprint
  {http://arxiv.org/abs/hep-th/9804132} {arXiv:hep-th/9804132} \BibitemShut
  {NoStop}%
\bibitem [{\citenamefont {Detar}\ and\ \citenamefont
  {Kogut}(1987)}]{Detar:1987kae}%
  \BibitemOpen
  \bibfield  {author} {\bibinfo {author} {\bibfnamefont {C.~E.}\ \bibnamefont
  {Detar}}\ and\ \bibinfo {author} {\bibfnamefont {J.~B.}\ \bibnamefont
  {Kogut}},\ }\href {\doibase 10.1103/PhysRevLett.59.399} {\bibfield  {journal}
  {\bibinfo  {journal} {Phys. Rev. Lett.}\ }\textbf {\bibinfo {volume} {59}},\
  \bibinfo {pages} {399} (\bibinfo {year} {1987})}\BibitemShut {NoStop}%
\bibitem [{\citenamefont {Bazavov}\ \emph
  {et~al.}(2019{\natexlab{b}})\citenamefont {Bazavov} \emph
  {et~al.}}]{Bazavov:2019www}%
  \BibitemOpen
  \bibfield  {author} {\bibinfo {author} {\bibfnamefont {A.}~\bibnamefont
  {Bazavov}} \emph {et~al.},\ }\href {\doibase 10.1103/PhysRevD.100.094510}
  {\bibfield  {journal} {\bibinfo  {journal} {Phys. Rev. D}\ }\textbf {\bibinfo
  {volume} {100}},\ \bibinfo {pages} {094510} (\bibinfo {year}
  {2019}{\natexlab{b}})},\ \Eprint {http://arxiv.org/abs/1908.09552}
  {arXiv:1908.09552 [hep-lat]} \BibitemShut {NoStop}%
\bibitem [{\citenamefont {Brandt}\ \emph {et~al.}(2016)\citenamefont {Brandt},
  \citenamefont {Francis}, \citenamefont {Meyer}, \citenamefont {Philipsen},
  \citenamefont {Robaina},\ and\ \citenamefont {Wittig}}]{Brandt:2016daq}%
  \BibitemOpen
  \bibfield  {author} {\bibinfo {author} {\bibfnamefont {B.~B.}\ \bibnamefont
  {Brandt}}, \bibinfo {author} {\bibfnamefont {A.}~\bibnamefont {Francis}},
  \bibinfo {author} {\bibfnamefont {H.~B.}\ \bibnamefont {Meyer}}, \bibinfo
  {author} {\bibfnamefont {O.}~\bibnamefont {Philipsen}}, \bibinfo {author}
  {\bibfnamefont {D.}~\bibnamefont {Robaina}}, \ and\ \bibinfo {author}
  {\bibfnamefont {H.}~\bibnamefont {Wittig}},\ }\href {\doibase
  10.1007/JHEP12(2016)158} {\bibfield  {journal} {\bibinfo  {journal} {JHEP}\
  }\textbf {\bibinfo {volume} {12}},\ \bibinfo {pages} {158} (\bibinfo {year}
  {2016})},\ \Eprint {http://arxiv.org/abs/1608.06882} {arXiv:1608.06882
  [hep-lat]} \BibitemShut {NoStop}%
\bibitem [{\citenamefont {Cheng}\ \emph {et~al.}(2011)\citenamefont {Cheng}
  \emph {et~al.}}]{Cheng:2010fe}%
  \BibitemOpen
  \bibfield  {author} {\bibinfo {author} {\bibfnamefont {M.}~\bibnamefont
  {Cheng}} \emph {et~al.},\ }\href {\doibase 10.1140/epjc/s10052-011-1564-y}
  {\bibfield  {journal} {\bibinfo  {journal} {Eur. Phys. J. C}\ }\textbf
  {\bibinfo {volume} {71}},\ \bibinfo {pages} {1564} (\bibinfo {year}
  {2011})},\ \Eprint {http://arxiv.org/abs/1010.1216} {arXiv:1010.1216
  [hep-lat]} \BibitemShut {NoStop}%
\bibitem [{\citenamefont {Dalla~Brida}\ \emph {et~al.}(2022)\citenamefont
  {Dalla~Brida}, \citenamefont {Giusti}, \citenamefont {Harris}, \citenamefont
  {Laudicina},\ and\ \citenamefont {Pepe}}]{DallaBrida:2021ddx}%
  \BibitemOpen
  \bibfield  {author} {\bibinfo {author} {\bibfnamefont {M.}~\bibnamefont
  {Dalla~Brida}}, \bibinfo {author} {\bibfnamefont {L.}~\bibnamefont {Giusti}},
  \bibinfo {author} {\bibfnamefont {T.}~\bibnamefont {Harris}}, \bibinfo
  {author} {\bibfnamefont {D.}~\bibnamefont {Laudicina}}, \ and\ \bibinfo
  {author} {\bibfnamefont {M.}~\bibnamefont {Pepe}},\ }\href {\doibase
  10.1007/JHEP04(2022)034} {\bibfield  {journal} {\bibinfo  {journal} {JHEP}\
  }\textbf {\bibinfo {volume} {04}},\ \bibinfo {pages} {034} (\bibinfo {year}
  {2022})},\ \Eprint {http://arxiv.org/abs/2112.05427} {arXiv:2112.05427
  [hep-lat]} \BibitemShut {NoStop}%
\bibitem [{\citenamefont {Eichten}\ and\ \citenamefont
  {Feinberg}(1981)}]{Eichten:1980mw}%
  \BibitemOpen
  \bibfield  {author} {\bibinfo {author} {\bibfnamefont {E.}~\bibnamefont
  {Eichten}}\ and\ \bibinfo {author} {\bibfnamefont {F.}~\bibnamefont
  {Feinberg}},\ }\href {\doibase 10.1103/PhysRevD.23.2724} {\bibfield
  {journal} {\bibinfo  {journal} {Phys. Rev. D}\ }\textbf {\bibinfo {volume}
  {23}},\ \bibinfo {pages} {2724} (\bibinfo {year} {1981})}\BibitemShut
  {NoStop}%
\bibitem [{\citenamefont {Koch}\ \emph {et~al.}(1992)\citenamefont {Koch},
  \citenamefont {Shuryak}, \citenamefont {Brown},\ and\ \citenamefont
  {Jackson}}]{Koch:1992nx}%
  \BibitemOpen
  \bibfield  {author} {\bibinfo {author} {\bibfnamefont {V.}~\bibnamefont
  {Koch}}, \bibinfo {author} {\bibfnamefont {E.~V.}\ \bibnamefont {Shuryak}},
  \bibinfo {author} {\bibfnamefont {G.~E.}\ \bibnamefont {Brown}}, \ and\
  \bibinfo {author} {\bibfnamefont {A.~D.}\ \bibnamefont {Jackson}},\ }\href
  {\doibase 10.1103/PhysRevD.46.3169} {\bibfield  {journal} {\bibinfo
  {journal} {Phys. Rev. D}\ }\textbf {\bibinfo {volume} {46}},\ \bibinfo
  {pages} {3169} (\bibinfo {year} {1992})},\ \bibinfo {note} {[Erratum:
  Phys.Rev.D 47, 2157 (1993)]},\ \Eprint {http://arxiv.org/abs/hep-ph/9204236}
  {arXiv:hep-ph/9204236} \BibitemShut {NoStop}%
\bibitem [{\citenamefont {Giusti}\ \emph {et~al.}(2001)\citenamefont {Giusti},
  \citenamefont {Paciello}, \citenamefont {Parrinello}, \citenamefont
  {Petrarca},\ and\ \citenamefont {Taglienti}}]{Giusti:2001xf}%
  \BibitemOpen
  \bibfield  {author} {\bibinfo {author} {\bibfnamefont {L.}~\bibnamefont
  {Giusti}}, \bibinfo {author} {\bibfnamefont {M.~L.}\ \bibnamefont
  {Paciello}}, \bibinfo {author} {\bibfnamefont {C.}~\bibnamefont
  {Parrinello}}, \bibinfo {author} {\bibfnamefont {S.}~\bibnamefont
  {Petrarca}}, \ and\ \bibinfo {author} {\bibfnamefont {B.}~\bibnamefont
  {Taglienti}},\ }\href {\doibase 10.1142/S0217751X01004281} {\bibfield
  {journal} {\bibinfo  {journal} {Int. J. Mod. Phys. A}\ }\textbf {\bibinfo
  {volume} {16}},\ \bibinfo {pages} {3487} (\bibinfo {year} {2001})},\ \Eprint
  {http://arxiv.org/abs/hep-lat/0104012} {arXiv:hep-lat/0104012} \BibitemShut
  {NoStop}%
\bibitem [{\citenamefont {Giusti}\ \emph {et~al.}(2024)\citenamefont {Giusti},
  \citenamefont {Harris}, \citenamefont {Laudicina}, \citenamefont {Pepe},\
  and\ \citenamefont {Rescigno}}]{Giusti:2024ohu}%
  \BibitemOpen
  \bibfield  {author} {\bibinfo {author} {\bibfnamefont {L.}~\bibnamefont
  {Giusti}}, \bibinfo {author} {\bibfnamefont {T.}~\bibnamefont {Harris}},
  \bibinfo {author} {\bibfnamefont {D.}~\bibnamefont {Laudicina}}, \bibinfo
  {author} {\bibfnamefont {M.}~\bibnamefont {Pepe}}, \ and\ \bibinfo {author}
  {\bibfnamefont {P.}~\bibnamefont {Rescigno}},\ }\href {\doibase
  10.1016/j.physletb.2024.138799} {\bibfield  {journal} {\bibinfo  {journal}
  {Phys. Lett. B}\ }\textbf {\bibinfo {volume} {855}},\ \bibinfo {pages}
  {138799} (\bibinfo {year} {2024})},\ \Eprint
  {http://arxiv.org/abs/2405.04182} {arXiv:2405.04182 [hep-lat]} \BibitemShut
  {NoStop}%
\bibitem [{\citenamefont {Mazur}\ \emph {et~al.}(2024)\citenamefont {Mazur}
  \emph {et~al.}}]{HotQCD:2023ghu}%
  \BibitemOpen
  \bibfield  {author} {\bibinfo {author} {\bibfnamefont {L.}~\bibnamefont
  {Mazur}} \emph {et~al.} (\bibinfo {collaboration} {HotQCD}),\ }\href
  {\doibase 10.1016/j.cpc.2024.109164} {\bibfield  {journal} {\bibinfo
  {journal} {Comput. Phys. Commun.}\ }\textbf {\bibinfo {volume} {300}},\
  \bibinfo {pages} {109164} (\bibinfo {year} {2024})},\ \Eprint
  {http://arxiv.org/abs/2306.01098} {arXiv:2306.01098 [hep-lat]} \BibitemShut
  {NoStop}%
\bibitem [{\citenamefont {Bazavov}\ \emph {et~al.}(2013)\citenamefont
  {Bazavov}, \citenamefont {Ding}, \citenamefont {Hegde}, \citenamefont
  {Karsch}, \citenamefont {Miao}, \citenamefont {Mukherjee}, \citenamefont
  {Petreczky}, \citenamefont {Schmidt},\ and\ \citenamefont
  {Velytsky}}]{Bazavov:2013uja}%
  \BibitemOpen
  \bibfield  {author} {\bibinfo {author} {\bibfnamefont {A.}~\bibnamefont
  {Bazavov}}, \bibinfo {author} {\bibfnamefont {H.~T.}\ \bibnamefont {Ding}},
  \bibinfo {author} {\bibfnamefont {P.}~\bibnamefont {Hegde}}, \bibinfo
  {author} {\bibfnamefont {F.}~\bibnamefont {Karsch}}, \bibinfo {author}
  {\bibfnamefont {C.}~\bibnamefont {Miao}}, \bibinfo {author} {\bibfnamefont
  {S.}~\bibnamefont {Mukherjee}}, \bibinfo {author} {\bibfnamefont
  {P.}~\bibnamefont {Petreczky}}, \bibinfo {author} {\bibfnamefont
  {C.}~\bibnamefont {Schmidt}}, \ and\ \bibinfo {author} {\bibfnamefont
  {A.}~\bibnamefont {Velytsky}},\ }\href {\doibase 10.1103/PhysRevD.88.094021}
  {\bibfield  {journal} {\bibinfo  {journal} {Phys. Rev. D}\ }\textbf {\bibinfo
  {volume} {88}},\ \bibinfo {pages} {094021} (\bibinfo {year} {2013})},\
  \Eprint {http://arxiv.org/abs/1309.2317} {arXiv:1309.2317 [hep-lat]}
  \BibitemShut {NoStop}%
\bibitem [{\citenamefont {Follana}\ \emph {et~al.}(2007)\citenamefont
  {Follana}, \citenamefont {Mason}, \citenamefont {Davies}, \citenamefont
  {Hornbostel}, \citenamefont {Lepage}, \citenamefont {Shigemitsu},
  \citenamefont {Trottier},\ and\ \citenamefont {Wong}}]{PhysRevD.75.054502}%
  \BibitemOpen
  \bibfield  {author} {\bibinfo {author} {\bibfnamefont {E.}~\bibnamefont
  {Follana}}, \bibinfo {author} {\bibfnamefont {Q.}~\bibnamefont {Mason}},
  \bibinfo {author} {\bibfnamefont {C.}~\bibnamefont {Davies}}, \bibinfo
  {author} {\bibfnamefont {K.}~\bibnamefont {Hornbostel}}, \bibinfo {author}
  {\bibfnamefont {G.~P.}\ \bibnamefont {Lepage}}, \bibinfo {author}
  {\bibfnamefont {J.}~\bibnamefont {Shigemitsu}}, \bibinfo {author}
  {\bibfnamefont {H.}~\bibnamefont {Trottier}}, \ and\ \bibinfo {author}
  {\bibfnamefont {K.}~\bibnamefont {Wong}},\ }\href {\doibase
  10.1103/PhysRevD.75.054502} {\bibfield  {journal} {\bibinfo  {journal} {Phys.
  Rev. D}\ }\textbf {\bibinfo {volume} {75}},\ \bibinfo {pages} {054502}
  (\bibinfo {year} {2007})}\BibitemShut {NoStop}%
\bibitem [{\citenamefont {Lepage}(1999)}]{PhysRevD.59.074502}%
  \BibitemOpen
  \bibfield  {author} {\bibinfo {author} {\bibfnamefont {G.~P.}\ \bibnamefont
  {Lepage}},\ }\href {\doibase 10.1103/PhysRevD.59.074502} {\bibfield
  {journal} {\bibinfo  {journal} {Phys. Rev. D}\ }\textbf {\bibinfo {volume}
  {59}},\ \bibinfo {pages} {074502} (\bibinfo {year} {1999})}\BibitemShut
  {NoStop}%
\bibitem [{\citenamefont {Bazavov}\ \emph {et~al.}(2010)\citenamefont
  {Bazavov}, \citenamefont {Bernard}, \citenamefont {DeTar}, \citenamefont
  {Freeman}, \citenamefont {Gottlieb}, \citenamefont {Heller}, \citenamefont
  {Hetrick}, \citenamefont {Laiho}, \citenamefont {Levkova}, \citenamefont
  {Oktay}, \citenamefont {Osborn}, \citenamefont {Sugar}, \citenamefont
  {Toussaint},\ and\ \citenamefont {Van~de Water}}]{PhysRevD.82.074501}%
  \BibitemOpen
  \bibfield  {author} {\bibinfo {author} {\bibfnamefont {A.}~\bibnamefont
  {Bazavov}}, \bibinfo {author} {\bibfnamefont {C.}~\bibnamefont {Bernard}},
  \bibinfo {author} {\bibfnamefont {C.}~\bibnamefont {DeTar}}, \bibinfo
  {author} {\bibfnamefont {W.}~\bibnamefont {Freeman}}, \bibinfo {author}
  {\bibfnamefont {S.}~\bibnamefont {Gottlieb}}, \bibinfo {author}
  {\bibfnamefont {U.~M.}\ \bibnamefont {Heller}}, \bibinfo {author}
  {\bibfnamefont {J.~E.}\ \bibnamefont {Hetrick}}, \bibinfo {author}
  {\bibfnamefont {J.}~\bibnamefont {Laiho}}, \bibinfo {author} {\bibfnamefont
  {L.}~\bibnamefont {Levkova}}, \bibinfo {author} {\bibfnamefont
  {M.}~\bibnamefont {Oktay}}, \bibinfo {author} {\bibfnamefont
  {J.}~\bibnamefont {Osborn}}, \bibinfo {author} {\bibfnamefont {R.~L.}\
  \bibnamefont {Sugar}}, \bibinfo {author} {\bibfnamefont {D.}~\bibnamefont
  {Toussaint}}, \ and\ \bibinfo {author} {\bibfnamefont {R.~S.}\ \bibnamefont
  {Van~de Water}} (\bibinfo {collaboration} {MILC Collaboration}),\ }\href
  {\doibase 10.1103/PhysRevD.82.074501} {\bibfield  {journal} {\bibinfo
  {journal} {Phys. Rev. D}\ }\textbf {\bibinfo {volume} {82}},\ \bibinfo
  {pages} {074501} (\bibinfo {year} {2010})}\BibitemShut {NoStop}%
\bibitem [{\citenamefont {Philipsen}(2002{\natexlab{a}})}]{Philipsen:2001ip}%
  \BibitemOpen
  \bibfield  {author} {\bibinfo {author} {\bibfnamefont {O.}~\bibnamefont
  {Philipsen}},\ }\href {\doibase 10.1016/S0550-3213(02)00089-5} {\bibfield
  {journal} {\bibinfo  {journal} {Nucl. Phys. B}\ }\textbf {\bibinfo {volume}
  {628}},\ \bibinfo {pages} {167} (\bibinfo {year} {2002}{\natexlab{a}})},\
  \Eprint {http://arxiv.org/abs/hep-lat/0112047} {arXiv:hep-lat/0112047}
  \BibitemShut {NoStop}%
\bibitem [{\citenamefont {Philipsen}(2002{\natexlab{b}})}]{Philipsen:2002az}%
  \BibitemOpen
  \bibfield  {author} {\bibinfo {author} {\bibfnamefont {O.}~\bibnamefont
  {Philipsen}},\ }\href {\doibase 10.1016/S0370-2693(02)01777-X} {\bibfield
  {journal} {\bibinfo  {journal} {Phys. Lett. B}\ }\textbf {\bibinfo {volume}
  {535}},\ \bibinfo {pages} {138} (\bibinfo {year} {2002}{\natexlab{b}})},\
  \Eprint {http://arxiv.org/abs/hep-lat/0203018} {arXiv:hep-lat/0203018}
  \BibitemShut {NoStop}%
\bibitem [{\citenamefont {Bazavov}\ \emph
  {et~al.}(2014{\natexlab{b}})\citenamefont {Bazavov}, \citenamefont
  {Bhattacharya}, \citenamefont {DeTar}, \citenamefont {Ding}, \citenamefont
  {Gottlieb}, \citenamefont {Gupta}, \citenamefont {Hegde}, \citenamefont
  {Heller}, \citenamefont {Karsch}, \citenamefont {Laermann}, \citenamefont
  {Levkova}, \citenamefont {Mukherjee}, \citenamefont {Petreczky},
  \citenamefont {Schmidt}, \citenamefont {Schroeder}, \citenamefont {Soltz},
  \citenamefont {Soeldner}, \citenamefont {Sugar}, \citenamefont {Wagner},\
  and\ \citenamefont {Vranas}}]{PhysRevD.90.094503}%
  \BibitemOpen
  \bibfield  {author} {\bibinfo {author} {\bibfnamefont {A.}~\bibnamefont
  {Bazavov}}, \bibinfo {author} {\bibfnamefont {T.}~\bibnamefont
  {Bhattacharya}}, \bibinfo {author} {\bibfnamefont {C.}~\bibnamefont {DeTar}},
  \bibinfo {author} {\bibfnamefont {H.-T.}\ \bibnamefont {Ding}}, \bibinfo
  {author} {\bibfnamefont {S.}~\bibnamefont {Gottlieb}}, \bibinfo {author}
  {\bibfnamefont {R.}~\bibnamefont {Gupta}}, \bibinfo {author} {\bibfnamefont
  {P.}~\bibnamefont {Hegde}}, \bibinfo {author} {\bibfnamefont {U.~M.}\
  \bibnamefont {Heller}}, \bibinfo {author} {\bibfnamefont {F.}~\bibnamefont
  {Karsch}}, \bibinfo {author} {\bibfnamefont {E.}~\bibnamefont {Laermann}},
  \bibinfo {author} {\bibfnamefont {L.}~\bibnamefont {Levkova}}, \bibinfo
  {author} {\bibfnamefont {S.}~\bibnamefont {Mukherjee}}, \bibinfo {author}
  {\bibfnamefont {P.}~\bibnamefont {Petreczky}}, \bibinfo {author}
  {\bibfnamefont {C.}~\bibnamefont {Schmidt}}, \bibinfo {author} {\bibfnamefont
  {C.}~\bibnamefont {Schroeder}}, \bibinfo {author} {\bibfnamefont {R.~A.}\
  \bibnamefont {Soltz}}, \bibinfo {author} {\bibfnamefont {W.}~\bibnamefont
  {Soeldner}}, \bibinfo {author} {\bibfnamefont {R.}~\bibnamefont {Sugar}},
  \bibinfo {author} {\bibfnamefont {M.}~\bibnamefont {Wagner}}, \ and\ \bibinfo
  {author} {\bibfnamefont {P.}~\bibnamefont {Vranas}} (\bibinfo {collaboration}
  {HotQCD Collaboration}),\ }\href {\doibase 10.1103/PhysRevD.90.094503}
  {\bibfield  {journal} {\bibinfo  {journal} {Phys. Rev. D}\ }\textbf {\bibinfo
  {volume} {90}},\ \bibinfo {pages} {094503} (\bibinfo {year}
  {2014}{\natexlab{b}})}\BibitemShut {NoStop}%
\bibitem [{\citenamefont {Bazavov}\ \emph
  {et~al.}(2012{\natexlab{b}})\citenamefont {Bazavov}, \citenamefont
  {Bhattacharya}, \citenamefont {Cheng}, \citenamefont {DeTar}, \citenamefont
  {Ding}, \citenamefont {Gottlieb}, \citenamefont {Gupta}, \citenamefont
  {Hegde}, \citenamefont {Heller}, \citenamefont {Karsch}, \citenamefont
  {Laermann}, \citenamefont {Levkova}, \citenamefont {Mukherjee}, \citenamefont
  {Petreczky}, \citenamefont {Schmidt}, \citenamefont {Soltz}, \citenamefont
  {Soeldner}, \citenamefont {Sugar}, \citenamefont {Toussaint}, \citenamefont
  {Unger},\ and\ \citenamefont {Vranas}}]{PhysRevD.85.054503}%
  \BibitemOpen
  \bibfield  {author} {\bibinfo {author} {\bibfnamefont {A.}~\bibnamefont
  {Bazavov}}, \bibinfo {author} {\bibfnamefont {T.}~\bibnamefont
  {Bhattacharya}}, \bibinfo {author} {\bibfnamefont {M.}~\bibnamefont {Cheng}},
  \bibinfo {author} {\bibfnamefont {C.}~\bibnamefont {DeTar}}, \bibinfo
  {author} {\bibfnamefont {H.-T.}\ \bibnamefont {Ding}}, \bibinfo {author}
  {\bibfnamefont {S.}~\bibnamefont {Gottlieb}}, \bibinfo {author}
  {\bibfnamefont {R.}~\bibnamefont {Gupta}}, \bibinfo {author} {\bibfnamefont
  {P.}~\bibnamefont {Hegde}}, \bibinfo {author} {\bibfnamefont {U.~M.}\
  \bibnamefont {Heller}}, \bibinfo {author} {\bibfnamefont {F.}~\bibnamefont
  {Karsch}}, \bibinfo {author} {\bibfnamefont {E.}~\bibnamefont {Laermann}},
  \bibinfo {author} {\bibfnamefont {L.}~\bibnamefont {Levkova}}, \bibinfo
  {author} {\bibfnamefont {S.}~\bibnamefont {Mukherjee}}, \bibinfo {author}
  {\bibfnamefont {P.}~\bibnamefont {Petreczky}}, \bibinfo {author}
  {\bibfnamefont {C.}~\bibnamefont {Schmidt}}, \bibinfo {author} {\bibfnamefont
  {R.~A.}\ \bibnamefont {Soltz}}, \bibinfo {author} {\bibfnamefont
  {W.}~\bibnamefont {Soeldner}}, \bibinfo {author} {\bibfnamefont
  {R.}~\bibnamefont {Sugar}}, \bibinfo {author} {\bibfnamefont
  {D.}~\bibnamefont {Toussaint}}, \bibinfo {author} {\bibfnamefont
  {W.}~\bibnamefont {Unger}}, \ and\ \bibinfo {author} {\bibfnamefont
  {P.}~\bibnamefont {Vranas}} (\bibinfo {collaboration} {HotQCD
  Collaboration}),\ }\href {\doibase 10.1103/PhysRevD.85.054503} {\bibfield
  {journal} {\bibinfo  {journal} {Phys. Rev. D}\ }\textbf {\bibinfo {volume}
  {85}},\ \bibinfo {pages} {054503} (\bibinfo {year}
  {2012}{\natexlab{b}})}\BibitemShut {NoStop}%
\bibitem [{\citenamefont {Giovannangeli}(2004)}]{Giovannangeli:2003ti}%
  \BibitemOpen
  \bibfield  {author} {\bibinfo {author} {\bibfnamefont {P.}~\bibnamefont
  {Giovannangeli}},\ }\href {\doibase 10.1016/j.physletb.2004.02.011}
  {\bibfield  {journal} {\bibinfo  {journal} {Phys. Lett. B}\ }\textbf
  {\bibinfo {volume} {585}},\ \bibinfo {pages} {144} (\bibinfo {year}
  {2004})},\ \Eprint {http://arxiv.org/abs/hep-ph/0312307}
  {arXiv:hep-ph/0312307} \BibitemShut {NoStop}%
\bibitem [{\citenamefont {Aoki}\ \emph {et~al.}(2022)\citenamefont {Aoki} \emph
  {et~al.}}]{FlavourLatticeAveragingGroupFLAG:2021npn}%
  \BibitemOpen
  \bibfield  {author} {\bibinfo {author} {\bibfnamefont {Y.}~\bibnamefont
  {Aoki}} \emph {et~al.} (\bibinfo {collaboration} {Flavour Lattice Averaging
  Group (FLAG)}),\ }\href {\doibase 10.1140/epjc/s10052-022-10536-1} {\bibfield
   {journal} {\bibinfo  {journal} {Eur. Phys. J. C}\ }\textbf {\bibinfo
  {volume} {82}},\ \bibinfo {pages} {869} (\bibinfo {year} {2022})},\ \Eprint
  {http://arxiv.org/abs/2111.09849} {arXiv:2111.09849 [hep-lat]} \BibitemShut
  {NoStop}%
\bibitem [{\citenamefont {Schmidt}\ and\ \citenamefont
  {Steinhauser}(2012)}]{Schmidt:2012az}%
  \BibitemOpen
  \bibfield  {author} {\bibinfo {author} {\bibfnamefont {B.}~\bibnamefont
  {Schmidt}}\ and\ \bibinfo {author} {\bibfnamefont {M.}~\bibnamefont
  {Steinhauser}},\ }\href {\doibase 10.1016/j.cpc.2012.03.023} {\bibfield
  {journal} {\bibinfo  {journal} {Comput. Phys. Commun.}\ }\textbf {\bibinfo
  {volume} {183}},\ \bibinfo {pages} {1845} (\bibinfo {year} {2012})},\ \Eprint
  {http://arxiv.org/abs/1201.6149} {arXiv:1201.6149 [hep-ph]} \BibitemShut
  {NoStop}%
\bibitem [{\citenamefont {Schroder}(1999)}]{Schroder:1999sg}%
  \BibitemOpen
  \bibfield  {author} {\bibinfo {author} {\bibfnamefont {Y.}~\bibnamefont
  {Schroder}},\ }\emph {\bibinfo {title} {{The Static potential in QCD}}},\
  \href {http://www-library.desy.de/cgi-bin/showprep.pl?desy-thesis99-021}
  {Ph.D. thesis},\ \bibinfo  {school} {Hamburg U.} (\bibinfo {year}
  {1999})\BibitemShut {NoStop}%
\end{thebibliography}%

\end{document}